\documentclass[showpacs,aps,prd,nofootinbib,floatfix,amsmath,amssymb]{revtex4}
\usepackage{graphicx}
\usepackage{slashed}
\usepackage{mathtools}
\usepackage{multirow,array}
\usepackage{tikz}
\usetikzlibrary{decorations.markings}
\numberwithin{equation}{section}
\begin{document}

\makeatletter
%Feynman slash
\newbox\slashbox \setbox\slashbox=\hbox{$/$}
\newbox\Slashbox \setbox\Slashbox=\hbox{\large$/$}
\def\pFMslash#1{\setbox\@tempboxa=\hbox{$#1$}
  \@tempdima=0.5\wd\slashbox \advance\@tempdima 0.5\wd\@tempboxa
  \copy\slashbox \kern-\@tempdima \box\@tempboxa}
\def\pFMSlash#1{\setbox\@tempboxa=\hbox{$#1$}
  \@tempdima=0.5\wd\Slashbox \advance\@tempdima 0.5\wd\@tempboxa
  \copy\Slashbox \kern-\@tempdima \box\@tempboxa}
\def\FMslash{\protect\pFMslash}
\def\FMSlash{\protect\pFMSlash}
\def\miss#1{\ifmmode{/\mkern-11mu #1}\else{${/\mkern-11mu #1}$}\fi}
%%%% Uso:  \pFMSlash{p}
\makeatother

%\tightenlines
\title{Implications of extra dimensions on the effective charge and the beta function in quantum electrodynamics}
\author{E. Mart\'\i nez-Pascual$ {}^{(a)}$}
\author{G. N\' apoles-Ca\~nedo${}^{(b)}$}
\author{H. Novales-S\' anchez${}^{(b)}$}
\author{A. Sierra-Mart\'\i nez${}^{(b)}$}
\author{J. J. Toscano${}^{(b)}$}
\address{$^{(a)}$ Departamento de Ciencias Naturales y Exactas,
Centro Universitario de los Valles, Universidad de Guadalajara,
Carretera Guadalajara-Ameca Km 45.5, CP 46000, Ameca, Jalisco, M\'exico.\\
$^{(b)}$Facultad de Ciencias F\'{\i}sico Matem\'aticas,
Benem\'erita Universidad Aut\'onoma de Puebla, Apartado Postal
1152, Puebla, Puebla, M\'exico.}
\begin{abstract}
A comprehensive analysis on the photon self-energy, the fermion self-energy, and the fermion vertex function is presented at one loop in the context of quantum electrodynamics (QED) with 1 extra dimension. In 5-dimensional theories, characterized by an infinite number of Kaluza-Klein fields, one-loop amplitudes involve discrete as well as continuous sums, $\sum^\infty_{n=1}\int d^4k$, that could diverge. Using dimensional regularization, we express such sums as products of gamma and Epstein functions, both defined on the complex plane, with divergences arising from poles of these functions in the limit as $ D \to 4$. Using the analytical properties of the Epstein function, we show that the ultraviolet divergences generated by the Kaluza-Klein sums can be consistently renormalized, which means that the corresponding renormalized quantities reduce to the usual ones of QED at the limit of a very large compactification scale $R^{-1}$. The main features of QED at the one-loop level were studied. We use the mass-dependent $\mu$-scheme to calculate, in QED with an arbitrary number $n$ of extra dimensions, a beta function fulfilling all desirable physical requirements. We argue that in this type of theories, with a large mass spectrum covering a wide energy range, beta functions should not be calculated by using mass-independent renormalization schemes. We show that the beta function is finite for any energy $\mu$. In particular, it reduces to the usual QED result $e^3/12\pi^2$ for $m\ll \mu \ll R^{-1}$ and vanishes for $m\gg \mu$, with $m$ the usual fermion mass. Throughout the work, the decoupling nature of all our results obtained from the analytical properties of the Epstein function is stressed.
\end{abstract}


\maketitle
\section{Introduction}
\label{I} Quantum field theories in more than 4 spacetime dimensions became phenomenologically attractive since Antoniadis, Arkani-Hamed, Dimopoulos, and Dvali~\cite{A,ADD,AADD} argued that relatively large extra dimensions could show up at the TeV scale. Shortly after, L.
Randall and R. Sundrum introduced the notion of warped extra dimensions to tackle the hierarchy problem~\cite{RS}. Another well-known extra-dimensional approach is the so-called universal extra dimensions (UED)~\cite{UED}, characterized by the assumption that all the dynamic variables propagate in the compact dimensions. In the UED framework, the starting point consists in formulating the standard model (SM) in a flat spacetime manifold ${\cal M}^{4+n}={\cal M}^4\times {\cal N}^n$, where ${\cal M}^4$ is the usual Minkowski space and ${\cal N}^n$ is a $n$-dimensional Euclidean manifold. In this stage, one assumes that distance scales in consideration are so small compared with the size of the extra dimensions that the SM in $4+n$ dimensions is correctly governed by the $(4+n)$-dimensional Poincar\' e group, ${\rm ISO}(1,3+n)$, and by the gauge group ${\rm SU}_C(3,{\cal M}^{4+n})\times {\rm SU}_L(2,{\cal M}^{4+n})\times{\rm U}_Y(1,{\cal M}^{4+n})$\footnote{This extension of the usual SM group differs only in the support spacetime manifold, meaning that ${\cal M}^{4+n}$ is used instead of ${\cal M}^4$.}. This is an effective field theory in $4+n$ spacetime dimensions with an infinite number of Lagrangian terms, which include a replica of the 4-dimensional SM and the set of all interactions of higher-than-4 mass dimensions. At lower energies, when the finite size of the manifold ${\cal N}^n$ is apparent, one needs to pass from the ${\rm ISO}(1,3+n)\times {\rm SU}_C(3,{\cal M}^{4+n})\times{\rm SU}_L(2,{\cal M}^{4+n})\times{\rm U}_Y(1,{\cal M}^{4+n})$ description, suitable for $4+n$ dimensions, to the standard 4-dimensional description, provided by ${\rm ISO}(1,3)\times{\rm SU}_C(3,{\cal M}^{4})\times{\rm SU}_L(2,{\cal M}^{4})\times{\rm U}_Y(1,{\cal M}^{4})$, which is achieved through an appropriate compactification scheme, followed by two canonical transformations~\cite{OP1,OP2,OP3} that allow us to map covariant objects of the extended groups into covariant objects of the standard groups. The process of hiding the extended symmetry into the standard symmetry leads to an effective theory in which each SM field has an associated infinite set of Kaluza-Klein (KK) modes, which are quantized in the standard way~\cite{OP3}. Symbolically, we start from a finite set of fields $\{\varphi_a(x,\bar x)\}$ (with $x\in {\cal M}^4, \bar x \in {\cal N}^n$ and $a$ a covariance index) governed by the extended groups, and then we pass to a set of fields that comprises the 4-dimensional SM fields $\{\varphi_a(x)\}$ and an infinite number of KK fields $\{ \varphi^{(\underline{n})}_a(x)\}$ (with $(\underline{n})$ a collection of natural indices) governed by the standard groups. Both descriptions are equivalent because one passes from one to the other through a canonical transformation. In this approach, conservation of extra-dimensional momentum leads to an effective theory that preserves KK parity, which introduces dynamical restrictions between SM and KK particles. In particular, KK effects on SM observables first arise at one loop, which show us the importance of studying the one-loop structure of this type of theories.\\

The phenomenological impact of 1 extra dimension on observables sensitive to new-physics effects has been a subject of interest in the literature. This is the case, for instance, of Higgs physics~\cite{HP}, flavor physics~\cite{OP2,FVP}, the electroweak gauge sector~\cite{GP}, $B$ physics~\cite{BMP}, and collider physics~\cite{CMS,ATLAS}. All these investigations show that the one-loop contributions of the infinite number of KK modes lead to amplitudes free of divergences. However, as far as we know, no investigations addressing contributions from 5-dimensional field formulations to amplitudes sensitive to ultraviolet (UV) divergences, such as self energies or vertex functions that require renormalization, exist. This sets an unusual challenge, as one must consider the UV divergencies from the one-loop contributions induced by an infinite number of fields, which may be a source of a new class of divergences. If the number of KK excitations were finite, no matter how large, we would be in a conventional scenario of calculating the one-loop contribution of a large, but finite, number of particles to a given vertex function. These types of scenarios are common in many extensions of the SM. Nonetheless, in our case, where the number of KK fields is infinite, the infinite sum involved in considering all these contributions may or may not converge. In this type of theories, a typical one-loop amplitude will involve, besides the usual continuous sum, a discrete infinite KK-mode sum, that is, $\sum^\infty_{n=1}\int d^4k$. To handle possible divergences, both continuous and discrete sums must be regularized. As it is well known, the dimensional-regularization scheme~\cite{RD1,RD2} has proven to be the best known tool for handling short-distance effects in the usual spacetime ${\cal M}^4$. In this scheme, the spacetime dimension is promoted to $D=4-\epsilon$ dimensions, being $\epsilon$ a complex number. Divergences, if they exist, appear as poles of the gamma function in the limit as $\epsilon \to 0$. In this paper, we show that this scheme can be used to simultaneously regularize both the discrete and continuous sums by using the analytical properties of the Epstein zeta function~\cite{E1}, which is a generalization of the Riemann zeta function~\cite{RF}. We will show that one-loop amplitudes can naturally be expressed as products of gamma functions and Epstein functions, with the divergencies from continuous and discrete sums appearing as the poles of the gamma function and the Epstein function, respectively. The main goal of this work is to develop this idea in the context of quantum electrodynamics (QED) with one UED, which we refer to by the acronym 5DQED. We focus on the one-loop impact of KK fields on the three basic Green functions of QED, namely, the fermion self-energy, the photon self-energy and the fermion vertex function. This means that we must consider the one-loop contributions of the infinite set of KK modes associated with the usual spinor field and the gauge field that describe some charged fermion and the photon in QED. One of the main objectives of this work is to show how this type of divergences can be regularized and consistently absorbed by the parameters of the theory. Although we will focus only in one extra dimension, some of the more relevant results will be discussed in the broader context of an arbitrary number of extra dimensions. In particular, we will address aspects of vacuum polarization.\\

One of the main goals of this work is to show how dimensional regularization allows us to control, through the gamma function  and the Epstein zeta function, the divergences that can arise from continuous and discrete sums, respectively. To show the internal consistence of our approach, we study many of the well-known one-loop properties of QED. Besides verification of the Ward identity at one loop and the correction to the anomalous magnetic dipole moment, we calculate the effective charge~\cite{EEC} in the presence of an arbitrary number $n$ of extra dimensions and study some of its more important implications at the one-loop level, such as gauge independence, the Thompson limit, and the correction to Coulomb's Law. The study of the beta function deserves special attention. We discuss the disadvantages of using a mass-independent renormalization scheme in the calculation of beta function in this type of theories. The reason is the presence of massive particles in a wide range of energies. So, we show that a physically acceptable beta function for this type of theories can be obtained using the mass-dependent scheme known as the $\mu$-scheme, with $\mu$ the subtraction point. We show that the beta function so calculated satisfies all physical requirements and reduces to the well-known value obtained in mass-independent schemes, such as MS or $\overline{\rm MS}$. The decoupling nature of new-physics effects arising from extra dimensions is shown to occur in all calculated one-loop amplitudes.\\

The paper has been organized as follows. The basic structure of 5DQED, including the Feynman rules needed for our calculations, will be discussed in Sec.\ref{5DQED}. The one-loop structure of the fermion self-energy, the photon self-energy, the fermion vertex function, and the Ward-Takahashi identity are studied in Sec.\ref{LOOP}. The impact of an arbitrary number $n$ of extra dimensions on the vacuum polarization is explored in Sec.\ref{VPn}.  Finally, we will present our conclusions in Sec.\ref{C}.

\section{QED with one extra dimension}
\label{5DQED}
First of all, we recall that there is no chirality in odd-dimensional spinor formulations. This means that field theories defined in spacetimes with an odd number of dimensions are necessarily vector-like. The construction of the 5-dimensional SM and, in particular, QED requires the symmetry dictated by the orbifold $S^1/Z_2$ (with $S^1$ the circle of radius $R$), used to dimensionally reduce the theory. In five dimensions, as in the 4-dimensional case, Dirac fields are still objects with four components. The corresponding generators are given by $S^{MN}=\frac{i}{4}[\gamma^M,\gamma^N]$, with $\gamma^M=\gamma^\mu, i\gamma^5$ the standard Dirac matrices, which satisfy the Clifford's algebra $\{\gamma^M, \gamma^N\}=2g^{MN}$. Throughout the paper we will use a metric with negative signature, that is,  $g={\rm diag}(+1,-1,-1, -1,-1)$.\\

The generation of the mass terms for zero modes in 5DQED is somewhat subtle. The problem has been addressed from two different, but equivalent, perspectives~\cite{OP2,PVSM,OP4}. In one of these approaches, we start from the fact that QED is embedded in the electroweak theory and then we generate the mass of the zero mode via the Higgs mechanism~\cite{OP2,OP4}. This leads to a doubly mass-degenerate KK spectrum $\psi^{(n)}_{(1)}$ and $\psi^{(n)}_{(2)}$ associated with the zero mode fermion field $\psi^{(0)}\equiv \psi$. The other approach consists in assuming QED as a self-contained theory, so the mass of the zero mode is generated by introducing a set of mirror fermions~\cite{PVSM}. In this case, the zero mode $\psi$ has associated a mass-degenerate double KK spectrum. Both approaches lead to the same dynamics of the charged fermion $\psi$, the electromagnetic gauge field $A^{(0)}_\mu\equiv A_\mu$, and their KK excitations $(\psi^{(n)}_{(1)}, \psi^{(n)}_{(2)})$ and $(A^{(n)}_\mu, A^{(n)}_{\rm G})$, respectively. A comprehensive analysis on the matter is given in Refs.~\cite{OP2,PVSM,OP4}, so we restrict ourselves to present those results that are needed four our purposes. In the case of only one extra dimension there are no physical scalar fields. The only scalar field is the pseudo-Goldstone boson $A^{(n)}_{\rm G}$ associated with the gauge field $A^{(n)}_\mu$ and emerged from the KK mass-generation mechanism.\\

The 4-dimensional effective KK Lagrangian is given by
\begin{equation}
{\cal L}_{\rm eff \, QED}={\cal L}_{\rm QED}+{\cal L}_{\text{0-KK}}+{\cal L}_{\rm KK}+{\cal L}_{\mathbf{d>4}}\, ,
\end{equation}
with ${\cal L}_{\rm QED}$ the standard QED Lagrangian, given by
\begin{equation}
\label{qed}
{\cal L}_{\rm QED}=\bar \psi(i\pFMSlash{D}-m)\psi-\frac{1}{4}F_{\mu \nu}F^{\mu \nu}-\frac{1}{\xi}(\partial_\mu A^\mu)^2\, ,
\end{equation}
where $\xi$ is the gauge-fixing parameter. The Lagrangian ${\cal L}_{\text{0-KK}}$ plays a central role in our study because it contains the interactions between KK zero modes and excited modes. As it was anticipated, the KK excitations of $\psi$, $\psi^{(n)}_{(1)}$ and  $\psi^{(n)}_{(2)}$, are mass degenerate, being their masses given by $m^2_{\psi^{(n)}}=m^2+m^2_{(n)}$, with $m^2_{(n)}\equiv p^{(n)}_5 p^{(n)}_5=(n/R)^2$. Here $p^{(n)}_5$ is the component of the momentum along the compact dimension. A mixing between $\psi^{(n)}_{(1)}$ and  $\psi^{(n)}_{(2)}$ arises, which is characterized by the angle
\begin{equation}
\label{angle}
\tan \theta_{\psi^{(n)}}=\sqrt{\frac{m_{\psi^{(n)}}+m_{(n)}}{m_{\psi^{(n)}}-m_{(n)}}}\, .
\end{equation}
The Lagrangian ${\cal L}_{\text{0-KK}}$ is given by
\begin{eqnarray}
\label{0kk}
{\cal L}_{\text{0-KK}}&=&\bar \psi(i\pFMSlash{D}-m)\psi+\sum^\infty_{n=1}\left[\bar \psi^{(n)}_{(1)}(i\pFMSlash{D}-m_{\psi^{(n)}})\psi^{(n)}_{(1)}+\bar{ \psi}^{(n)}_{(2)}(i\pFMSlash{D}-m_{\psi^{(n)}}) {\psi}^{(n)}_{(2)}\right]\nonumber \\
&&+Q_{\psi}e\sum^\infty_{n=1}A^{(n)}_\mu \Big[\bar \psi \gamma^\mu \left(s_{\psi^{(n)}}P_L+c_{\psi^{(n)}}P_R \right)\psi^{(n)}_{(1)}+
\bar \psi^{(n)}_{(1)} \gamma^\mu \left(s_{\psi^{(n)}}P_L+c_{\psi^{(n)}}P_R \right)\psi\nonumber \\
&&-\bar \psi \gamma^\mu \left(c_{\psi^{(n)}}P_L+s_{\psi^{(n)}}P_R \right)\psi^{(n)}_{(2)}-
\bar \psi^{(n)}_{(2)} \gamma^\mu \left(c_{\psi^{(n)}}P_L+s_{\psi^{(n)}}P_R \right)\psi\Big]\nonumber \\
&&+iQ_{\psi}e\sum^\infty_{n=1}A^{(n)}_G\Big[\bar \psi\left(s_{\psi^{(n)}}P_R-c_{\psi^{(n)}}P_L\right)\psi^{(n)}_{(1)}
-\bar \psi^{(n)}_{(1)}\left(s_{\psi^{(n)}}P_L-c_{\psi^{(n)}}P_R\right)\psi \nonumber \\
&&+\bar \psi\left(c_{\psi^{(n)}}P_R-s_{\psi^{(n)}}P_L\right)\psi^{(n)}_{(2)}
-\bar \psi^{(n)}_{(2)}\left(c_{\psi^{(n)}}P_L-s_{\psi^{(n)}}P_R\right)\psi\Big]\, ,
\end{eqnarray}
where $s_{\psi^{(n)}}$ and  $c_{\psi^{(n)}}$ stand for the sine and the cosine of the mixing angle $\theta_{\psi^{(n)}}$. On the other hand, ${\cal L}_{\rm KK}$ represents interactions only among KK excitations. In this case, we only display the quadratic parts, which is needed to define the propagators of the KK excitations of the electromagnetic gauge field. Then, we have
\begin{eqnarray}
\label{kk}
{\cal L}_{\rm KK}&=&- \frac{1}{4}\sum^\infty_{n=1}F^{(n)}_{\mu \nu}F^{(n)\mu \nu}+\sum^\infty_{n=1}\left[\frac{1}{2}
(\partial_\mu A^{(n)}_G)(\partial^\mu A^{(n)}_G)+m_{(n)}A^{(n)}_\mu (\partial^\mu A^{(n)}_G)+\frac{1}{2}A^{(n)}_\mu A^{(n)\mu} \right]\nonumber \\
&&-\sum^\infty_{n=1} \frac{1}{2\xi_{(n)}}f^{(n)}f^{(n)}+\cdots
\end{eqnarray}
where the last term, involving products $f^{(n)}f^{(n)}$, represents the gauge-fixing term for gauge invariance characterized by the KK gauge parameters $\alpha^{(n)}$\footnote{The five-dimensional gauge parameter $\alpha(x,\bar x)$ is assumed to be even under $\bar x \to -\bar x$, so it is Fourier expanded as $\alpha(x,\bar x)=\frac{1}{\sqrt{2\pi R}} \alpha^{(0)}(x)+\sum^\infty_{n=1}\frac{1}{\sqrt{\pi R}}\cos\left(\frac{n\bar x}{R}\right) \alpha^{(n)}(x)$. The zero mode is identified as the standard gauge parameter $\alpha\equiv \alpha^{(0)}$.}. The gauge-fixing functions for excited gauge modes are given by $f^{(n)}=\partial_\mu A^{(n)\mu}-\xi_{(n)}m_{(n)}A^{(n)}_G$, with $\xi_{(n)}$ the gauge-fixing parameter.  We stress that the Lagrangians displayed in Eqs.~(\ref{qed}), (\ref{0kk}), and (\ref{kk}) contain only renormalizable interactions in the sense that their canonical dimension is less than or equal to 4. This is so because these interactions arise from the compactification of the 5-dimensional version of QED. This fact has important consequences at the one-loop level.\\

Finally, ${\cal L}_{\mathbf{d>4}}$ contains all the interactions of canonical dimension higher than 4 that are compatible with the ${\rm ISO}(1,3)\times{\rm U}_Q(1,{\cal M}^4)$ symmetry. This type of interactions must be present in the 4-dimensional KK theory because the 5-dimensional theory is nonrenormalizable, according to power-counting criterion (Dyson's criterion), and thus must already be present even before compactification. In this work we will not consider loop insertions coming from this sector.\\

\begin{figure}
\centering
\includegraphics[width=10cm]{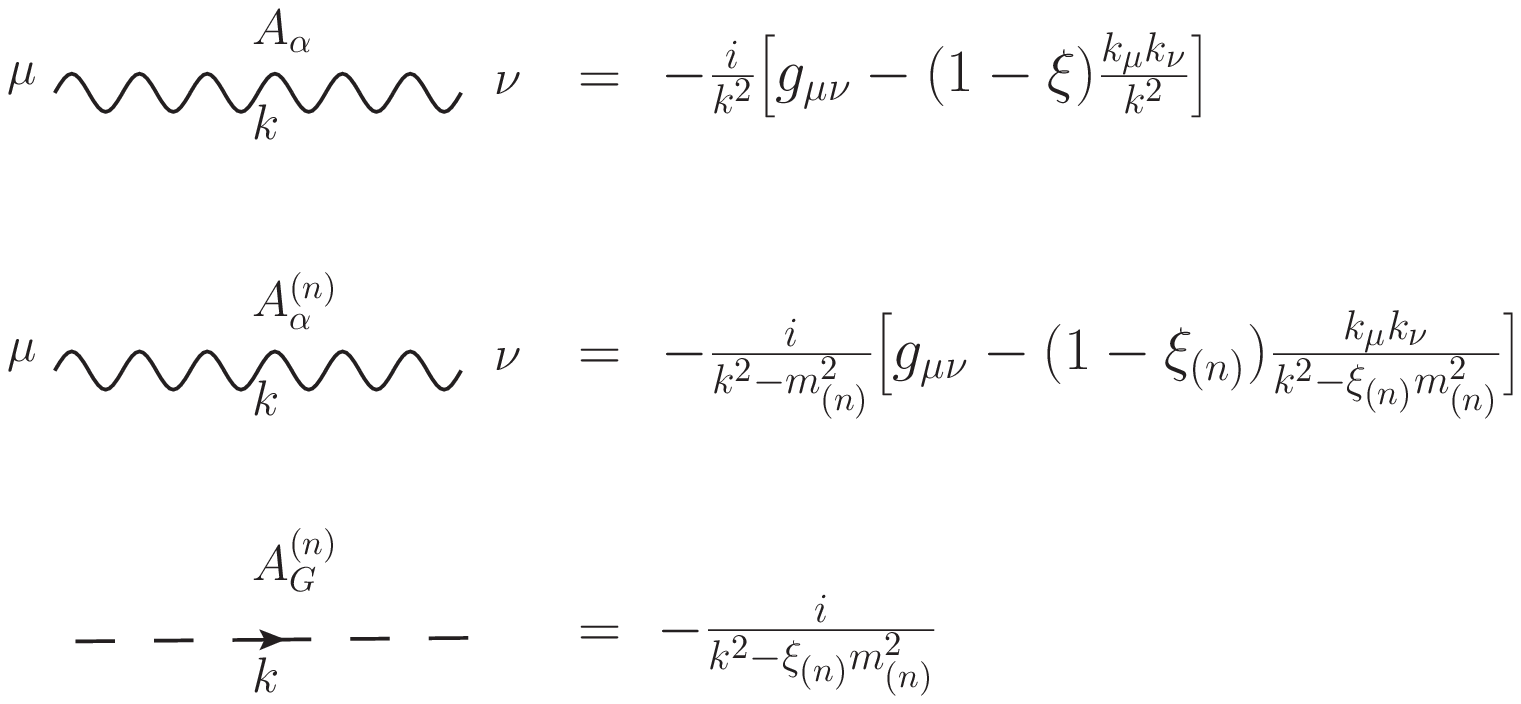}
\caption{\label{P} {\footnotesize Free propagators in the $R_\xi$-gauge. The calculations are performed in the Feynman-'t Hooft gauge ($\xi=1$ and $\xi_{(n)}=1$).}}
\end{figure}

\begin{figure}
\centering
\includegraphics[width=16cm]{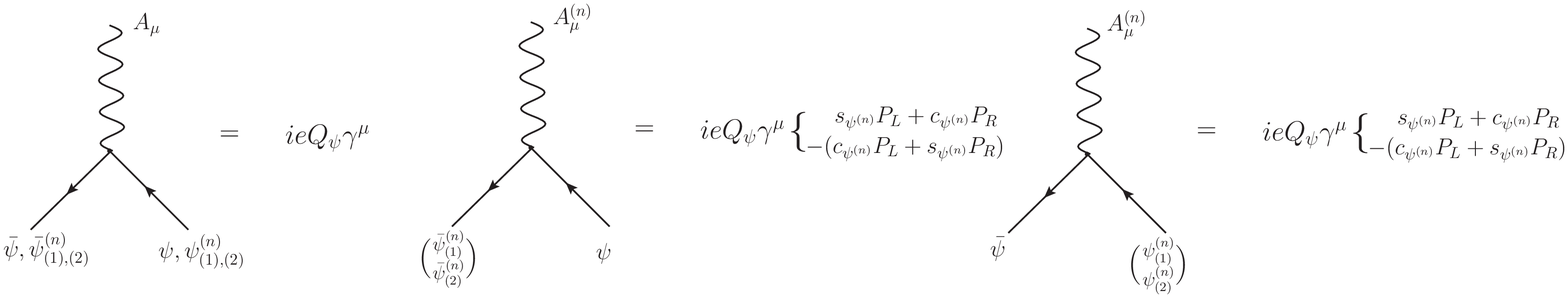}
\vspace{0.2cm} \\
\includegraphics[width=16cm]{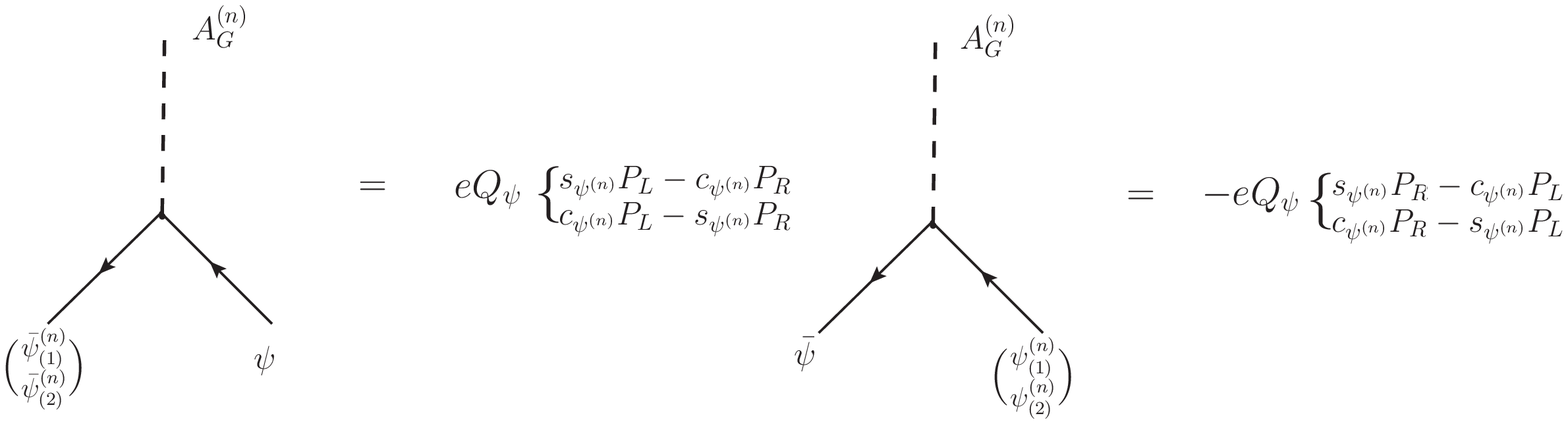}
\caption{\label{VFR} {\footnotesize Vertices needed for the calculation of self-energies and vertex function in 5DQED.}}
\end{figure}

The Feynman rules needed for our calculations are shown, in the $R_\xi$-gauge, in Figs.~\ref{P} and \ref{VFR}.

\section{One-loop effects of one extra dimension in QED}

\label{LOOP} In this section, we study the one-loop structure of the photon self-energy, the fermion self-energy, and the vertex function $\bar \psi \psi \gamma$, in the context of QED with one extra dimension.\\

As usual, we relate renormalized quantities $\{\psi, A_\mu, e \}$ with bare quantities $\{\psi_B, A_{\mu B}, e_B \}$ through renormalization factors as follows:
\begin{equation}
\label{Re}
\psi_B=\sqrt{Z_2}\psi\, , \, \, \, \,  A_{\mu B}=\sqrt{Z_3} A_\mu \, , \, \, \, \,  e_B=\left(\frac{Z_1}{Z_2}\right)Z^{-\frac{1}{2}}_3 e \, .
\end{equation}
Although we will not consider vertex functions containing external excited legs, we also define the corresponding relations for the KK excited  modes. In this case, we have
\begin{equation}
 \psi^{(n)}_{(1),(2)\, B}=\sqrt{Z^{(n)}_{(1),(2)\, 2}}\psi^{(n)}\, , \, \, \, \,   A^{(n)}_{\mu B}=\sqrt{Z^{(n)}_3} A^{(\underline{m})}_\mu , \, \, \, \,  A^{(n)}_{G\, B}=\sqrt{Z^{(n)}_G} A^{(n)}_{G}\, .
\end{equation}
Then, the bare Lagrangian can be written as
\begin{equation}
\label {ELR5}
{\cal L}^{\rm bare}_{\rm eff \, QED}={\cal L}_{\rm QED}+{\cal L}_{\text{0-KK}}+{\cal L}_{\rm KK}+{\cal L}_{\mathbf{d>4}}+{\cal L}^{\rm QED}_{\rm c.t.}+{\cal L}^{\rm KK}_{\rm c.t.}+{\cal L}^{\textbf{d}>4}_{\rm c.t}\, ,
\end{equation}
where ${\cal L}_{\rm QED}$, ${\cal L}_{\text{0-KK}}$, and ${\cal L}_{\rm KK}$ represent the renormalized Lagrangians given by Eqs.(\ref{qed}), (\ref{0kk}), and (\ref{kk}), respectively, while ${\cal L}^{\textbf{d}>4}$ contains the interactions of dimension higher than four written in terms of renormalized quantities. In addition, ${\cal L}^ {\rm QED}_{\rm c.t.}$ represents the standard or usual counterterm of QED, which is given by
\begin{equation}
\label{CT5}
{\cal L}^{\rm QED}_{\rm c.t.}=-\frac{1}{4}\delta_3F_{\mu \nu}F^{\mu \nu}+\bar{\psi}\left(i\delta_2\pFMSlash{\partial} -\delta_m\right)\psi+eQ_{\psi}\delta_1 \bar \psi \gamma^\mu \psi A_\mu \, ,
\end{equation}
where
\begin{equation}
\label{Rd}
\delta_3=Z_3-1\, , \, \, \, \, \delta_2=Z_2-1\, , \, \, \, \, \delta_1=Z_1-1\, , \, \, \, \, \delta_m=Z_2 m_B-m\, .
\end{equation}
In Eq.~(\ref{ELR5}), ${\cal L}^{\rm KK}_{\rm c.t.}$ and ${\cal L}^{\textbf{d}>4}_{\rm c.t}$ represent the counterterms that contain interactions between  standard and KK fields, which we do not present in this section, since we will not need them.\\

\subsection{Fermion self-energy}
\label{FSE} 
\begin{figure}
\centering
\includegraphics[width=12cm]{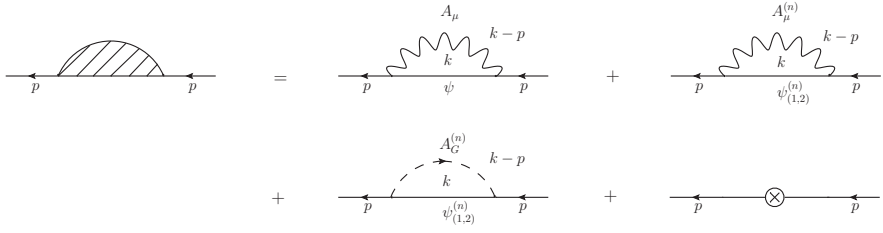}
\caption{\label{FS} {\footnotesize } Feynman diagrams contributing to the fermion self-energy in the Feynman-t'Hooft gauge. The counterterm diagram is also shown. }
\end{figure}

The one-loop contribution to the fermion self-energy is given by the Feynman diagrams shown in Fig.~\ref{FS}. The renormalized self-energy can be written as follows:
\begin{equation}
-i\Sigma_{\rm 5D}(p)=-i\Sigma(p)-i\Sigma_{\rm KK}(p)-i\Sigma_{\rm c.t.}(p)\, ,
\end{equation}
where $-i\Sigma(p)$, $-i\Sigma_{\rm KK}(p)$, and $-i\Sigma_{\rm c.t.}(p)$ represent the standard one-loop contribution, the one-loop contribution of the infinite number of KK modes, and the standard counterterm, respectively. In the Feynman-'t Hooft gauge ($\xi=1$ and $\xi_{(n)}=1$), the Feynman rules given in Figs.~\ref{P} and \ref{VFR} lead to the following amplitudes:
\begin{subequations}
\begin{align}
\label{SFSE}
-i\Sigma(p)&=-e^2Q^2_\psi(\mu^2)^{2-\frac{D}{2}} \int \frac{d^Dk}{(2\pi)^D}\frac{\gamma_\mu (\pFMSlash{k}+m)\gamma^\mu}{[k^2-m^2][(k-p)^2-m^2_\gamma]}\, ,\\
\label{KKFSE}
-i\Sigma_{KK}(p)&=-e^2Q^2_{\psi}(\mu^2)^{2-\frac{D}{2}}\sum^\infty_{n=1}\int \frac{d^Dk}{(2\pi)^D}\frac{T_A+T_G}{[k^2-m^2_{\psi^{(n)}}][(k-p)^2-m^2_{(n)}]}\, ,\\
\label{CTFSE}
-i\Sigma_{c.t.}(p)&=i(\pFMSlash{p}\delta_2-\delta_m)\, ,
\end{align}
\end{subequations}
where $\mu$ is the dimensional-regularization scale. In the standard contribution, Eq.~(\ref{SFSE}), we have regularized the infrared divergence by adding a small photon mass $m_{\gamma}$. In addition, the terms $T_A$ and $T_G$ in (\ref{KKFSE}) stand for the contributions of the gauge field $A^{(n)}_\mu$ and its pseudo-Goldstone boson $A^{(n)}_{\rm G}$, respectively. They are given by
\begin{subequations}
\begin{align}
T_A&=\gamma_\mu \left(\pFMSlash{k}+m\right)\gamma^\mu \, , \\
T_G&=-\pFMSlash{k}+m\, ,
\end{align}
\end{subequations}
where in obtaining these results we have used the relation $2s_{\psi^{(n)}}c_{\psi^{(n)}}m_{\psi^{(n)}}=m$. \\

Using Feynman parametrization and shifting $k\to k+xp$, the standard and KK contributions can be written as
\begin{subequations}
\begin{align}
\Sigma(p)&=\frac{\alpha Q^2_\psi}{4\pi}\int^1_0dx(4\pi \mu^2)^{2-\frac{D}{2}}\frac{1}{i\pi^{\frac{D}{2}}}\int d^Dk \frac{-(D-2)x\pFMSlash{p}+Dm}{\left(k^2-\hat{\Delta}^2_{2F}\right)^2}\, ,\\
\Sigma_{\rm KK}(p)&=\frac{\alpha Q^2_\psi}{4\pi}\int^1_0dx(4\pi \mu^2)^{2-\frac{D}{2}}\sum^\infty_{n=1}\frac{1}{i\pi^{\frac{D}{2}}}\int d^Dk \frac{-(D-1)x\pFMSlash{p}+(D+1)m}{\left(k^2-\Delta^2_{(n)F}\right)^2}\, ,
\end{align}
\end{subequations}
 where $\Delta^2_{(n)F}=m^2_{(n)}+\Delta^2_{2F}$, with $\Delta^2_{2F}=(1-x)m^2-x(1-x)p^2$, and $\hat{\Delta}^2_{2F}=\Delta^2_{2F}+xm^2_\gamma$. Note that the KK contribution is free of infrared divergences. After solving the integrals on $k$, we obtain
 \begin{equation}
 \Sigma(p)=\frac{\alpha Q^2_\psi}{4\pi}\int^1_0dx f(p)\, \Gamma\left(\frac{\epsilon}{2}\right)\left(\frac{\hat{\Delta}^2_{2F}}{4\pi \mu^2}\right)^{-\frac{\epsilon}{2}}\, ,
 \end{equation}
 for the standard contribution, where $f(p)=-(2-\epsilon)x\pFMSlash{p}+(4-\epsilon)m$. As far as the KK contribution is concerned, we have
 \begin{eqnarray}
 \Sigma_{\rm KK}(p)&=&\frac{\alpha Q^2_\psi}{4\pi}\int^1_0dx \, g(p)\, \Gamma\left(\frac{\epsilon}{2}\right)\sum^\infty_{n=1}\left(\frac{\Delta^2_{(n)F}}{4\pi \mu^2}\right)^{-\frac{\epsilon}{2}}\nonumber \\
 &=&\frac{\alpha Q^2_\psi}{4\pi}\int^1_0dx \, g(p)\, \Gamma\left(\frac{\epsilon}{2}\right)\left(\frac{R^{-2}}{4\pi \mu^2}\right)^{-\frac{\epsilon}{2}}
 \sum^\infty_{n=1}\frac{1}{(n^2+c^2_{2F})^{\frac{\epsilon}{2}}}\nonumber \\
 &=&\frac{\alpha Q^2_\psi}{4\pi}\int^1_0dx \, g(p)\, \Gamma\left(\frac{\epsilon}{2}\right)\left(\frac{R^{-2}}{4\pi \mu^2}\right)^{-\frac{\epsilon}{2}}
 E^{c^2_{2F}}_1\left(\frac{\epsilon}{2}\right)\, ,
 \end{eqnarray}
 where $g(p)=-(3-\epsilon)x\pFMSlash{p}+(5-\epsilon)m$. In the last step we have introduced the one-dimensional Epstein zeta function, which is defined as
 \begin{equation}
 E^{c^2}_1(s)=\sum^\infty_{n=1}\frac{1}{(n^2+c^2)^{s}}\, .
 \end{equation}
 This function is a generalization of the Riemann zeta function,
 \begin{equation}
 \zeta(s)=\sum^\infty_{n=1}\frac{1}{n^s}\, .
 \end{equation}
In our case, $c^2_{2F}=\frac{\Delta^2_{2F}}{R^{-2}}$ and $s=\epsilon/2$. Note that both the Gamma function and the Epstein function are defined on the  complex plane. Since the one-dimensional Epstein function has simple poles at $s=\frac{1}{2},-\frac{1}{2}, -\frac{3}{2},\cdots$~\cite{K}, it is clear that $E^{c^2_{2F}}_1\left(\frac{\epsilon}{2}\right)$ converges in the $\epsilon \to 0$ limit. This is a remarkable result, which emerges as a consequence of the analytical properties of the Epstein function.\\

From the above results, the one-loop contribution to the self-energy of the fermion $\psi$ can be written as follows:
\begin{equation}
\Sigma_{\rm 5D}(p)=\frac{\alpha Q^2_\psi}{4\pi}\int^1_0dx\, \Gamma\left(\frac{\epsilon}{2}\right)\left[f(p)\left(\frac{\hat{\Delta}^2_{2F}}{4\pi \mu^2}\right)^{-\frac{\epsilon}{2}}+g(p)\left(\frac{R^{-2}}{4\pi \mu^2}\right)^{-\frac{\epsilon}{2}}E^{c^2_{2F}}_1\left(\frac{\epsilon}{2}\right) \right]+\delta_m-\pFMSlash{p}\delta_2\, .
\end{equation}
To determine the counterterms $\delta_m$ and $\delta_2$ we use the on-shell renormalization conditions
\begin{subequations}
\begin{align}
\Sigma_{\rm 5D}(\pFMSlash{p})\Big|_{\pFMSlash{p}=m}&=0\, , \\
\frac{d}{d\pFMSlash{p}}\Sigma_{\rm 5D}(\pFMSlash{p})\Big|_{\pFMSlash{p}=m}&=0\, .
\end{align}
\end{subequations}
These renormalization conditions lead to
\begin{eqnarray}
\delta_2&=&\frac{\alpha Q^2_\psi}{4\pi}\int^1_0dx\, \Gamma\left(\frac{\epsilon}{2}\right)\Bigg\{\left(\frac{\bar{\hat{\Delta}}^2_{2F}}{4\pi \mu^2}\right)^{-\frac{\epsilon}{2}}\left[\frac{d\bar{f}}{d\pFMSlash{p}}-\frac{\epsilon}{2}\left(\frac{2m\bar{f}}{\bar{\hat{\Delta}}^2_{2F}}\right)
\left(\frac{d\bar{\hat{\Delta}}^2_{2F}}{dp^2}\right)\right]\nonumber \\
&&+\left(\frac{R^{-2}}{4\pi \mu^2}\right)^{-\frac{\epsilon}{2}}\left[\frac{d\bar{g}}{d\pFMSlash{p}}E^{\bar{c}^2_{2F}}_1\left(\frac{\epsilon}{2}\right)-
\frac{\epsilon}{2} \left(2m\bar{g}\right) E^{\bar{c}^2_{2F}}_1\left(1+\frac{\epsilon}{2}\right)\left(\frac{d\bar{c}^2_{2F}}{dp^2}\right)\right]
 \Bigg\}\, ,
\end{eqnarray}
\begin{eqnarray}
\delta_m&=&\frac{\alpha Q^2_\psi}{4\pi}\int^1_0dx\, \Gamma\left(\frac{\epsilon}{2}\right)\Bigg\{ \left(\frac{\bar{\hat{\Delta}}^2_{2F}}{4\pi \mu^2}\right)^{-\frac{\epsilon}{2}}\left[m\frac{d\bar{f}}{d\pFMSlash{p}}-\bar{f}-\frac{\epsilon}{2}\left(\frac{2m^2\bar{f}}{\bar{\hat{\Delta}}^2_{2F}}\right)
\left(\frac{d\bar{\hat{\Delta}}^2_{2F}}{dp^2}\right)\right]\nonumber \\
&&+\left(\frac{R^{-2}}{4\pi \mu^2}\right)^{-\frac{\epsilon}{2}}\left[\left(m\frac{d\bar{g}}{d\pFMSlash{p}}-\bar{g}\right)E^{\bar{c}^2_{2F}}_1\left(\frac{\epsilon}{2}\right)
-\frac{\epsilon}{2}\left(2m^2\bar{g}\right) E^{\bar{c}^2_{2F}}_1\left(1+\frac{\epsilon}{2}\right)\left(\frac{d\bar{c}^2_{2F}}{dp^2}\right)\right]
\Bigg\}\, ,
\end{eqnarray}
where we use the bar notation to indicate that the function under consideration has been evaluated at $\pFMSlash{p}= m$, that is, $\bar F=F(\pFMSlash{p}=m)$. In addition, we have used the chain rule
\begin{equation}
\frac{d}{d\pFMSlash{p}}=2\pFMSlash{p}\frac{d}{dp^2}=2\pFMSlash{p}\frac{dc^2_{2F}}{dp^2}\frac{d}{dc^2_{2F}}\, ,
\end{equation}
together with the fact that
\begin{equation}
\frac{dE^{c^2_{2F}}_1\left(\frac{\epsilon}{2}\right)}{dc^2_{2F}}=-\frac{\epsilon}{2}E^{c^2_{2F}}_1\left(1+\frac{\epsilon}{2}\right)\, .
\end{equation}
It is worthwhile to write the divergent parts of counterterms $\delta_2$ and $\delta_m$, separating explicitly the contributions arising from both the zero mode and the KK excitations. Keeping only the part proportional to the pole of the gamma function, we have:
\begin{subequations}
\begin{align}
\delta_2&=-\frac{\alpha Q^2_\psi}{4\pi}\left[\underbrace{\left(\frac{2}{\epsilon}\right)}_{\rm SC}+
\underbrace{\frac{3}{2}\zeta(0)\left(\frac{2}{\epsilon}\right)}_{\rm KKC}+\cdots\right]\, , \\
\delta_m&=-\frac{\alpha Q^2_\psi}{4\pi}\, m\left[\underbrace{4\left(\frac{2}{\epsilon}\right)}_{\rm SC}+
\underbrace{5\zeta(0)\left(\frac{2}{\epsilon}\right)}_{\rm KKC}+\cdots \right]\, ,
\end{align}
\end{subequations}
where SC and KKC are acronyms for ``standard contribution'' and ``KK contribution'', respectively. Note that $\zeta(0)=-1/2$. The importance of simultaneously regularizing discrete and continuous sums using the dimensional regularization scheme can now be appreciated.\\

On the other hand, using the relations
\begin{subequations}
\begin{align}
f(p)+(m-\pFMSlash{p})\frac{d\bar{f}}{d\pFMSlash{p}}-\bar{f}&=0\, , \\
g(p)+(m-\pFMSlash{p})\frac{d\bar{g}}{d\pFMSlash{p}}-\bar{g}&=0\, ,
\end{align}
\end{subequations}
we can write the renormalized fermion self-energy as
\begin{eqnarray}
\Sigma_{\rm 5D}(p)&=&\frac{\alpha Q^2_\psi}{4\pi}\int^1_0dx\, \Gamma\left(\frac{\epsilon}{2}\right)\Bigg\{ f(p)\left[ \left(\frac{\hat{\Delta}^2_{2F}}{4\pi \mu^2}\right)^{-\frac{\epsilon}{2}}- \left(\frac{\bar{\hat{\Delta}}^2_{2F}}{4\pi \mu^2}\right)^{-\frac{\epsilon}{2}}\right]\nonumber \\
&&+\left(\frac{R^{-2}}{4\pi \mu^2}\right)^{-\frac{\epsilon}{2}}g(p)\left[ E^{c^2_{2F}}_1\left(\frac{\epsilon}{2}\right) - E^{\bar{c}^2_{2F}}_1\left(\frac{\epsilon}{2}\right) \right]\nonumber \\
&&-\frac{\epsilon}{2}(m-\pFMSlash{p})\left[\left(\frac{2m\bar{f}}{\bar{\hat{\Delta}}^2_{2F}}\right)
\left(\frac{\bar{\hat{\Delta}}^2_{2F}}{dp^2}\right)+\left(\frac{R^{-2}}{4\pi \mu^2}\right)^{-\frac{\epsilon}{2}} \left(2m\bar{g}\right) E^{\bar{c}^2_{2F}}_1\left(1+\frac{\epsilon}{2}\right)\left(\frac{d\bar{c}^2_{2F}}{dp^2}\right)\right]
\Bigg\}\, .
\end{eqnarray}
In the above expressions, $\bar{\hat{\Delta}}^2_{2F}=(1-x)^2m^2+xm^2_\gamma$ and $\bar{c}^2_{2F}=(1-x)^2m^2/R^{-2}$. In addition, $\bar{f}=2(2-x)m$ and $\bar{g}=(5-3x)m$. Now, the following relations:
\begin{subequations}
\begin{align}
\label{IG}
\Gamma(s)&=\frac{1}{s}-\gamma_E+O(s)\, , \\
\label{IX}
X^{-s}&=1-s\log(X)+O(s^2)\, , \\
\label{IE}
E^{c^2}_1(s)&=E^{c^2}_1(0)+\frac{dE^{c^2}_1(s)}{ds}\Big|_{s=0}s+O(s^2)\nonumber \\
&=\zeta(0)-s\sum^\infty_{n=1}\log(n^2+c^2)+O(s^2)\, ,
\end{align}
\end{subequations}
valid near $s=0$, allow us to express the renormalized fermion self-energy as
\begin{eqnarray}
\label{FSE}
\Sigma_{\rm 5D}(p)&=&\frac{\alpha Q^2_\psi}{4\pi}\int^1_0dx\,\Bigg\{2(2m-x\pFMSlash{p})\log\left(\frac{\bar{\hat{\Delta}}^2_{2F}}{\hat{\Delta}^2_{2F}}\right)+
(m-\pFMSlash{p})\left[\frac{4x(1-x)(2-x)m^2}{(1-x)^2m^2+xm^2_\gamma}\right]\nonumber \\
&&+(5m-3x \pFMSlash{p})\sum^\infty_{n=1}\log\left(\frac{\bar{\Delta}^2_{(n)F}}{\Delta^2_{(n)F}}\right)+
(m-\pFMSlash{p})\left[2x(1-x)(5-3x)\left(\frac{m}{R^{-1}}\right)^2E^{\bar{c}^2_{2F}}_1(1)\right] \Bigg\}\, ,
\end{eqnarray}
with the first line of the above expression corresponding to the standard contribution, and the second one corresponding to the KK contribution. It is easy to show that effects from the extra dimension decouple in the limit as $R^{-1}\to \infty$. On the other hand, the multi-dimensional Epstein function can be expressed in power series of $c^2$~\cite{E2,E3}, which in the case of $E^{\bar{c}^2_{2F}}_1(1)$ becomes
\begin{equation}
\label{Ap1}
E^{\bar{c}^2_{2F}}_1(1)=\zeta(2)+F(\bar{c}^2_{2F})\, ,
\end{equation}
where
\begin{equation}
\label{Ap2}
F(\bar{c}^2_{2F})=\sum^\infty_{k=1}(-1)^k\zeta(2k+2)\bar{c}^{2k}_{2F}\, .
\end{equation}
Note that $\zeta(2)=\frac{\pi^2}{6}$.

\subsection{Photon self-energy}
\label{PSE} 
\begin{figure}
\centering
\includegraphics[width=12cm]{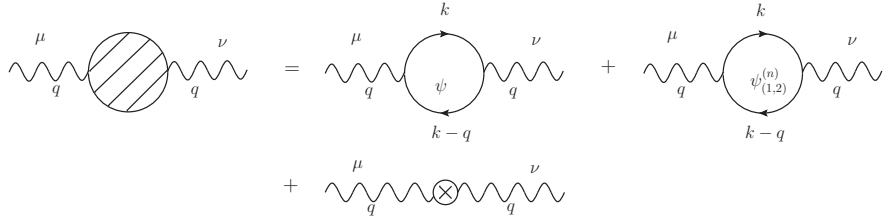}
\caption{\label{PS} {\footnotesize Feynman diagrams contributing to the the photon self-energy. The counterterm diagram also is included.}}
\end{figure}

We now proceed to calculate the one-loop contribution from 5DQED to the photon self-energy. This contribution is generated by the diagrams displayed in Fig.~\ref{PS}. Due to gauge invariance, the amplitude must have the following gauge structure: $i\Pi^{\mu \nu}(q)=i(q^2g^{\mu \nu}-q^\mu q^\nu)\Pi(q^2)$. The one-loop contribution, including the counterterm, is given by
\begin{equation}
i\Pi^{\mu \nu}_{\rm 5D}(q)=i\Pi^{\mu \nu}(q)+i\Pi^{\mu \nu}_{\rm KK}(q)+i\Pi^{\mu \nu}_{\rm c.t.}(q)\, ,
\end{equation}
with
\begin{subequations}
\begin{align}
i\Pi^{\mu \nu}(q)&=-e^2Q^2_\psi(\mu^2)^{2-\frac{D}{2}}\int \frac{d^Dk}{(2\pi)^D}\frac{tr\left[\gamma^\mu(\pFMSlash{k}+m)\gamma^\nu(\pFMSlash{k}+\pFMSlash{q}+m)\right]}{[k^2-m^2][(k+q)^2-m^2]}\, ,\\
i\Pi^{\mu \nu}_{\rm KK}(q)&=-e^2Q^2_\psi(\mu^2)^{2-\frac{D}{2}}\, 2\sum^\infty_{n=1}\int \frac{d^Dk}{(2\pi)^D}\frac{tr\left[\gamma^\mu(\pFMSlash{k}+m_{\psi^{(n)}})\gamma^\nu(\pFMSlash{k}+\pFMSlash{q}+m_{\psi^{(n)}})\right]}
{[k^2-m_{\psi^{(n)}}^2][(k+q)^2-m_{\psi^{(n)}}^2]}\, ,\\
i\Pi^{\mu \nu}_{\rm c.t.}(q)&=-i(q^2g^{\mu \nu}-q^\mu q^\nu)\delta_3\, .
\end{align}
\end{subequations}
In the KK contribution, $i\Pi^{\mu\nu}_{\rm KK}(q)$, a factor 2 has been introduced to take into account the contribution from the mass-degenerate pair of KK fermions $\psi^{(n)}_{(1)}$ and $\psi^{(n)}_{(2)}$. \\

Once Feynman parametrization has been implemented, we introduce the change of variables $k\to k-xq$, and then use the symmetry relation $k^\mu k^\nu \to k^2g^{\mu \nu}/D$ to obtain
\begin{subequations}
\begin{align}
i\Pi^{\mu \nu}(q)&=-4e^2Q^2_\psi(\mu^2)^{2-\frac{D}{2}}\int^1_0 dx  \int \frac{d^Dk}{(2\pi)^D}\frac{\left[\frac{2}{D}\left(1-\frac{D}{2}\right)k^2+
m^2+x(1-x)q^2\right]g^{\mu \nu}-2x(1-x)q^\mu q^\nu}{\left(k^2-\Delta^2_{2P}\right)^2}\, , \\
i\Pi^{\mu \nu}_{\rm KK}(q)&=-4e^2Q^2_\psi(\mu^2)^{2-\frac{D}{2}}\, 2\sum^\infty_{n=1}\int^1_0 dx  \int \frac{d^Dk}{(2\pi)^D}\frac{\left[\frac{2}{D}\left(1-\frac{D}{2}\right)k^2+
m^2_{\psi^{(n)}}+x(1-x)q^2\right]g^{\mu \nu}-2x(1-x)q^\mu q^\nu}{\left(k^2-\Delta^2_{(n)P}\right)^2}\, ,
\end{align}
\end{subequations}
where $\Delta^2_{2P}=m^2-x(1-x)q^2$ and $\Delta^2_{(n)P}=m^2_{(n)}+\Delta^2_{2P}$. Due to gauge invariance, there is no quadratic divergence, as the integral on $k^2$ is proportional to $(1-\frac{D}{2})\Gamma(1-\frac{D}{2})=\Gamma(2-\frac{D}{2})$. For the same reason, the dependence of both numerators on the mass disappears. Consequently, we can write the one-loop polarization functions as
\begin{subequations}
\begin{align}
\label{P1}
\Pi(q^2)&=-\frac{\alpha Q^2_\psi}{4\pi}\int^1_0dx\,f_P(x)\Gamma\left(\frac{\epsilon}{2}\right)
\left(\frac{\Delta^2_{2P}}{2\pi\mu^2}\right)^{-\frac{\epsilon}{2}}\, , \\
\Pi_{\rm KK}(q^2)&=-\frac{\alpha Q^2_\psi}{4\pi}\int^1_0dx\,f_P(x)\, 2\, \Gamma\left(\frac{\epsilon}{2}\right)\, \sum^\infty_{n=1}
\left(\frac{\Delta^2_{(n)P}}{4\pi\mu^2}\right)^{-\frac{\epsilon}{2}}\nonumber \\
\label{P2}
&=-\frac{\alpha Q^2_\psi}{4\pi}\int^1_0dx\,f_P(x)\left(\frac{R^{-2}}{4\pi\mu^2}\right)^{-\frac{\epsilon}{2}}\, 2\,
\Gamma\left(\frac{\epsilon}{2}\right)E^{c^2_{2P}}_1\left(\frac{\epsilon}{2}\right)\, ,
\end{align}
\end{subequations}
where $f_P(x)=8x(1-x)$ and $c^2_{2P}=\Delta^2_{2P}/R^{-2}$. Then, the renormalized  polarization function can be written as
\begin{equation}
\Pi_{\rm 5D}(q^2)=-\frac{\alpha Q^2_\psi}{4\pi}\int^1_0dx\,f_P(x)\Gamma\left(\frac{\epsilon}{2}\right)\left[\left(\frac{\Delta^2_{2P}}{2\pi\mu^2}\right)^{-\frac{\epsilon}{2}}
+\left(\frac{R^{-2}}{4\pi\mu^2}\right)^{-\frac{\epsilon}{2}}\, 2\,E^{c^2_{2P}}_1\left(\frac{\epsilon}{2}\right) \right]-\delta_3\, .
\end{equation}
To determine the counterterm, we impose the renormalization condition
\begin{equation}
\label{RCP}
\Pi_{\rm 5D}(0)=0\, ,
\end{equation}
which leads to
\begin{equation}
\delta_3=-\frac{\alpha Q^2_\psi}{4\pi}\int^1_0dx\,f_P(x)\Gamma\left(\frac{\epsilon}{2}\right)\left[\left(\frac{\bar{\Delta}^2_{2P}}{2\pi\mu^2}\right)^{-\frac{\epsilon}{2}}
+\left(\frac{R^{-2}}{4\pi\mu^2}\right)^{-\frac{\epsilon}{2}}\, 2\,E^{\bar{c}^2_{2P}}_1\left(\frac{\epsilon}{2}\right) \right]\, ,
\end{equation}
 where $\bar{\Delta}^2_{2P}=m^2$ and $\bar{c}^2_{2P}=m^2/R^{-2}$. As in the case of the fermion self-energy, it is worth analyzing the divergent part of this counterterm. Keeping only the pole of the gamma function, we have
\begin{equation}
\label{PDiv}
\delta_3=-\frac{\alpha Q^2_\psi}{3\pi}\left[\underbrace{\left(\frac{2}{\epsilon}\right)}_{\rm SC}+\underbrace{2\zeta(0)\left(\frac{2}{\epsilon}\right)}_{\rm KKC}+
\cdots \right]\, .
\end{equation}
Since $\zeta(0)=-1/2$, the KK divergence exactly cancels the standard divergence. This cancelation of the UV divergences at the one-loop level is a consequence of the double multiplicity of the KK excitations of the fermion $\psi$. However, as we will see below, this cancelation no longer occurs when there is more than one extra dimension.\\

Using the relations  given by Eqs.~(\ref{IG}), (\ref{IX}), and (\ref{IE}), the polarization function becomes
\begin{equation}
\label{VP}
\Pi_{\rm 5D}(q^2)=-\frac{\alpha Q^2_\psi}{4\pi}\int^1_0dx\,f_P(x)\left\{\log\left[\frac{m^2-x(1-x)q^2}{m^2}\right]+
2\sum^\infty_{n=1}\log\left[\frac{m^2_{\psi^{(n)}}-x(1-x)q^2}{m^2_{\psi^{(n)}}}\right]\right\}\, .
\end{equation}
The structure of this result is quite suggestive and reflects the consistency of our regularization scheme. In the first term of the above  expression we can recognize the usual contribution of QED to the vacuum polarization~\cite{Peskin}, while in the following term a replica of this contribution for each KK excitation appears. While the standard contribution arises essentially from terms of the form $\Gamma\left(\frac{\epsilon}{2}\right)
\left(\frac{\Delta^2_{2P}}{2\pi\mu^2}\right)^{-\frac{\epsilon}{2}}$, the KK contribution comes from terms of the form $\Gamma\left(\frac{\epsilon}{2}\right)E^{c^2_{2P}}_1\left(\frac{\epsilon}{2}\right)$ [see Eqs.(\ref{IG}), (\ref{IX}), and (\ref{IE})]. Note that the KK contributions vanishes in the $R^{-1}\to \infty$ limit.

\subsection{The vertex}
\label{V} 
\begin{figure}
\centering
\includegraphics[width=12cm]{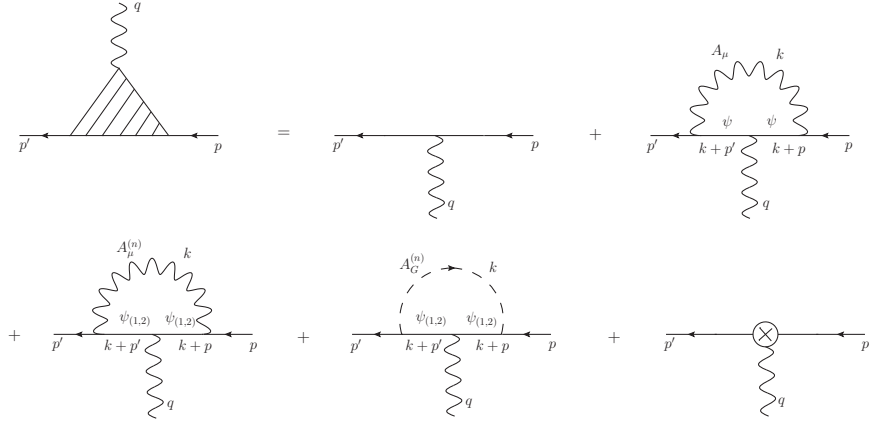}
\caption{\label{FV} {\footnotesize Feynman diagrams contribution to the fermion vertex function in the Feynman-t'Hooft gauge. The counterterm diagram also is displayed.}}
\end{figure}

Up to one-loop level, the contribution of 5DQED to the fermion vertex function $\bar\psi \psi \gamma$ is given by the Feynman diagrams shown in Fig.~\ref{FV}. The corresponding vertex function can be written as follows:
\begin{equation}
ieQ_\psi \Gamma^\mu_{\rm 5D}(p',p)=ieQ_\psi \gamma^\mu +ieQ_\psi\Gamma^\mu(p',p)+ieQ_\psi\Gamma^\mu_{\rm KK}(p',p)+ieQ_\psi\Gamma^\mu_{\rm c.t.}(p',p)\, ,
\end{equation}
where
\begin{subequations}
\begin{align}
\label{V1}
\Gamma^\mu(p',p)&=-ie^2Q^2_\psi (\mu^2)^{2-\frac{D}{2}}\int \frac{d^Dk}{(2\pi)^D}\frac{T^\mu}{[k^2-m^2_\gamma][(k+p)^2-m^2][(k+p')^2-m^2]}\, ,\\
\label{V2}
\Gamma^\mu_{\rm KK}(p',p)&=-ie^2Q^2_\psi (\mu^2)^{2-\frac{D}{2}}\sum^\infty_{n=1}\int \frac{d^Dk}{(2\pi)^D}\frac{T^\mu_A+T^\mu_G}{[k^2-m^2_{(n)}][(k+p)^2-m^2_{\psi^{(n)}}][(k+p')^2-m^2_{\psi^{(n)}}]}\, ,\\
\Gamma^\mu_{\rm c.t.}(p',p)&=\delta_1 \gamma^\mu\, ,
\end{align}
\end{subequations}
with
\begin{subequations}
\begin{align}
T^\mu&=(2-D)\pFMSlash{k}\gamma^\mu\pFMSlash{k}+4m(k+p+p')^\mu -4m^2\gamma^\mu \, , \\
T^\mu_A&=(2-D)\pFMSlash{k}\gamma^\mu \pFMSlash{k}+4m(k+p+p')^\mu-2(m^2+m^2_{\psi^{(n)}})\gamma^\mu \, , \\
T^\mu_G&=-\pFMSlash{k}\gamma^\mu \pFMSlash{k}-2(p+p')^\mu \pFMSlash{k}+\left[2(p+p')\cdot k+m^2-q^2-m^2_{\psi^{(n)}} \right]\gamma^\mu \, ,
\end{align}
\end{subequations}
with $T^\mu_A$ and $T^\mu_G$ representing the contributions, in the Feynman-'t Hooft gauge, of the gauge field $A^{(n)}_\mu$ and its pseudo-Goldstone boson $A^{(n)}_{\rm G}$, respectively. In obtaining the above results, the relation $m_{\psi^{(n)}}(2s_{\psi^{(n)}}c_{\psi^{(n)}})=m$ was used. \\

Using Feynman parametrization, Eqs.~(\ref{V1}) and (\ref{V2}) become
\begin{subequations}
\begin{align}
\label{V3}
\Gamma^\mu(p',p)&=-ie^2Q^2_\psi (\mu^2)^{2-\frac{D}{2}}I_P\Gamma(3)\int \frac{d^Dk}{(2\pi)^D}\frac{\hat{T}^\mu }{\left[k^2-\hat{\Delta}^2_{3V}\right]^3}\, ,\\
\label{V4}
\Gamma^\mu_{\rm KK}(p',p)&=-ie^2Q^2_\psi (\mu^2)^{2-\frac{D}{2}}I_P\Gamma(3)\sum^\infty_{n=1}\int \frac{d^Dk}{(2\pi)^D}\frac{\hat{T}^\mu_A+\hat{T}^\mu_G}{\left[k^2-\Delta^2_{(n)V}\right]^3}\, ,
\end{align}
\end{subequations}
where
\begin{equation}
I_P=\int^1_0 dx\,dy\,dz\,\delta(x+y+z-1)\, .
\end{equation}
In addition, $\hat{\Delta}^2_{3V}=zm^2_\gamma+ \Delta^2_{3V}$, where $\Delta^2_{3V}=-xyq^2+(1-z)^2m^2$ and $\Delta^2_{(n)V}=m^2_{(n)}+\Delta^2_{3V}$. In the above expressions,
\begin{subequations}
\begin{align}
\label{V5}
\hat{T}^\mu&=\left[\frac{(2-D)^2}{D}k^2 -2(1-4z+z^2)m^2-2(1-x)(1-y)q^2\right]\gamma^\mu -\left[4m^2z(1-z)\right]\frac{i\sigma^{\mu \nu}q_\nu}{2m}\, ,\\
\label{V6}
\hat{T}^\mu_A&=\left[\frac{(2-D)^2}{D}k^2 -2(-4z+z^2)m^2-2m^2_{\psi^{(n)}}-2(1-x)(1-y)q^2\right]\gamma^\mu -\left[4m^2z(1-z)\right]\frac{i\sigma^{\mu \nu}q_\nu}{2m}\, ,\\
\label{V7}
\hat{T}^\mu_G&=\left[-\frac{(2-D)}{D}k^2 +z(2-z)m^2-m^2_{\psi^{(n)}}-xyq^2\right]\gamma^\mu +\left[2m^2(1-z)^2\right]\frac{i\sigma^{\mu \nu}q_\nu}{2m}\, ,
\end{align}
\end{subequations}
where the Gordon identity has been used to eliminate $(p'+p)^\mu$ in favor of $i\sigma^{\mu \nu}q_\nu$. It is convenient to write these results in terms of electromagnetic form factors, that is,
\begin{equation}
\Gamma^\mu_{\rm 5D}(p',p)=F^{\rm 5D}_1(q^2)\gamma^\mu+F^{\rm 5D}_2(q^2)\frac{i\sigma^{\mu \nu}q_\nu}{2m}\, ,
\end{equation}
where
\begin{subequations}
\begin{align}
\label{5DFF1}
F^{\rm 5D}_1(q^2)&=1+F_1(q^2)+F^{\rm KK}_1(q^2)+\delta_1\, , \\
\label{5DFF2}
F^{\rm 5D}_2(q^2)&=F_2(q^2)+F^{\rm KK}_2(q^2)\, .
\end{align}
\end{subequations}
From Eqs.(\ref{V3})-(\ref{V7}), the one-loop contribution of 5DQED to the form factors $F^{\rm 5D}_1(q^2)$ and $F^{\rm 5D}_2(q^2)$ can be written as follows:
\begin{subequations}
\begin{align}
\label{SFF1}
F_1(q^2)&=\frac{\alpha Q^2_{\psi}}{4\pi}I_P\left[2\left(1-\frac{\epsilon}{2}\right)^2
\Gamma\left(\frac{\epsilon}{2}\right)\left(\frac{\hat{\Delta}^2_{3V}}{4\pi \mu^2}\right)^{-\frac{\epsilon}{2}}+\frac{f_V}{\hat{\Delta}^2_{3V}}\right]\, ,\\
\label{SFF2}
F_2(q^2)&=\frac{\alpha Q^2_{\psi}}{4\pi}I_P\frac{4m^2z(1-z)}{
\hat{\Delta}^2_{3V}}\, ,
\end{align}
\end{subequations}
\begin{subequations}
\begin{align}
\label{KKFF1}
F^{\rm KK}_1(q^2)&=\frac{\alpha Q^2_{\psi}}{4\pi}I_P\left[3\left(1-\frac{\epsilon}{2}\right)\left(1-\frac{\epsilon}{3}\right)\Gamma\left(\frac{\epsilon}{2}\right)
\left(\frac{R^{-2}}{4\pi \mu^2}\right)^{-\frac{\epsilon}{2}}E^{c^2_{3V}}_1\left(\frac{\epsilon}{2}\right)-\frac{\hat{f}}{R^{-2}}\, E^{c^2_{3V}}_1\left(1\right)+3\zeta(0)\right]\, , \\
\label{KKFF2}
F^{\rm KK}_2(q^2)&=\frac{\alpha Q^2_{\psi}}{4\pi}I_P\, 2(1-4z+3z^2)\left(\frac{m}{R^{-1}}\right)^2 E^{c^2_{3V}}_1\left(1\right)\, .
\end{align}
\end{subequations}
In the above expressions,
\begin{subequations}
\begin{align}
f_V&=2(1-4z+z^2)m^2+2(1-x)(1-y)q^2\, , \\
\hat{f}&=4zm^2-2(z+3xy)q^2\, .
\end{align}
\end{subequations}

To determine the counterterm, we impose the on-shell renormalization condition
\begin{equation}
\Gamma^\mu_{\rm 5D}(0)=\gamma^\mu\, ,
\end{equation}
which leads to
\begin{eqnarray}
\delta_1&=&-F_1(0)-F^{\rm KK}_1(0)\nonumber \\
&=&-\frac{\alpha Q^2_\psi}{4\pi}I_P\Bigg\{ \Gamma\left(\frac{\epsilon}{2}\right)\left[2\left(1-\frac{\epsilon}{2}\right)^2\left(\frac{\bar{\hat{\Delta}}^2_{3V}}{4\pi \mu^2}\right)^{-\frac{\epsilon}{2}}+3\left(1-\frac{\epsilon}{2}\right)\left(1-\frac{\epsilon}{3}\right)\left(\frac{R^{-2}}{4\pi \mu^2}\right)^{-\frac{\epsilon}{2}}E^{\bar{c}^2_{3V}}_1\left(\frac{\epsilon}{2}\right)\right]\nonumber \\
&&+\frac{\bar{f}_V}{\bar{\hat{\Delta}}^2_{3V}}-\frac{\bar{\hat{f}}}{R^{-2}}\, E^{\bar{c}^2_{3V}}_1\left(1\right)+3\zeta(0)
\Bigg\}\, ,
\end{eqnarray}
where $\bar{\Delta}^2_{3V}=\Delta^2_{3V}(q^2=0)=(1-z)^2m^2$,  $\bar{f}_V=f_V(q^2=0)=2(1-4z+z^2)m^2$, and $\bar{\hat{f}}=\hat{f}(q^2=0)=4zm^2$. The divergent part of this counterterm is given by
\begin{equation}
\delta_1=-\frac{\alpha Q^2_\psi}{4\pi}\left[\underbrace{\left(\frac{2}{\epsilon}\right)}_{\rm SC}+\underbrace{\frac{3}{2}\zeta(0)\left(\frac{2}{\epsilon}\right)}_{\rm KKC}+
\cdots \right]\, .
\end{equation}
For comparison purposes, it is convenient to explicitly write the renormalized form factor,
\begin{eqnarray}
\label{FF1}
F^{\rm 5D}_1(q^2)&=&1+\frac{\alpha Q^2_\psi}{4\pi}\int^1_0 dx\,dy\,dz\,\delta(x+y+z-1)\Bigg\{2\log\left(\frac{\bar{\hat{\Delta}}^2_{3V}}{\hat{\Delta}^2_{3V}}\right)+\frac{f_V}{\hat{\Delta}^2_{3V}}
-\frac{\bar{f}_V}{\bar{\hat{\Delta}}^2_{3V}}\nonumber \\
&&+3\sum^\infty_{n=1}\log\left(\frac{m^2_{(n)}+\bar{\Delta}^2_{3V}}{m^2_{(n)}+\Delta^2_{3V}}\right)+
\left(\frac{\bar{\hat{f}}}{R^{-2}}\right)E^{\bar{c}^2_{3V}}_1(1)-\left(\frac{\hat{f}}{R^{-2}}\right)E^{c^2_{3V}}_1(1)
\Bigg\}\, ,
\end{eqnarray}
where the first line corresponds to the standard contribution, while the second one is the KK contribution. As in the cases of the fermion self-energy and of the photon self-energy, it is easy to show that the KK contribution disappears from the expression (\ref{FF1}) in the limit as $R^{-1} \to \infty$.\\

As far as the form factor $F^{\rm KK}_2(q^2)$ is concerned, it can be written as follows:
\begin{equation}
F^{\rm KK}_2(q^2)=\frac{\alpha Q^2_\psi}{4\pi}\int^1_0 dx\,dy\,dz\,\delta(x+y+z-1)\left[\frac{4z(1-z)m^2}{\hat{\Delta}^2_{3V}}+
\frac{2(1-4z+3z^2)m^2}{R^{-2}}E^{c^2_{3V}}_1(1)\right]\, .
\end{equation}
As it occurs in QED, this form factor is free of both ultraviolet and infrared divergences, so $F^{\rm KK}_2(0)$ is a physical quantity, given by
\begin{eqnarray}
F^{\rm KK}_2(0)&=&\frac{\alpha Q^2_\psi}{4\pi}\int^1_0 dz(1-z)\left[\frac{4z}{1-z}+
2(1-4z+3z^2)\frac{m^2}{R^{-2}}E^{\bar{c}^2_{3V}}_1(1)\right]\nonumber \\
&=&\frac{\alpha Q^2_\psi}{4\pi}\left[2+f(mR)\right]\, ,
\end{eqnarray}
where
\begin{equation}
f(mR)=\sum^\infty_{k=0}(-1)^k\left(\frac{2k+1}{2k^2+7k+6}\right)\zeta(2k+2)(mR)^{2k+2}\, .
\end{equation}
Then, the correction to the $g$-factor of the fermion $\psi$ in 5DQED is:
\begin{equation}
a^{\rm 5D}_\psi=a_\psi\left[1+\frac{1}{2}f(mR)\right]\, ,
\end{equation}
where $a_\psi=\frac{\alpha Q^2_\psi}{2\pi}$ is the standard $g$-factor. Note that the KK contribution decouples from the QED prediction as powers of $mR$. The impact of extra dimensions on the muon anomalous magnetic moment has already been studied in the context of the SM~\cite{AMMM}.\\

\subsection{The Ward-Takahashi Identity}
\label{WI}
In QED, the fermion vertex function fulfills the Ward-Takahashi identity $q_\mu \Gamma^\mu=S^{-1}(p')-S^{-1}(p)$, with $S^{-1}$ the inverse of the fermion propagator. This identity is fulfilled to all orders of perturbation theory and guarantees that $F_1(0)=1$. At one loop, this identity implies that $\delta_1=\delta_2$ or $Z_1=Z_2$. Now we proceed to prove that this occurs in 5DQED as well. Using the relations given by Eqs.(\ref{IG})-(\ref{IE}), implementing an integration by parts to remove the standard contribution, and carrying out some algebra, we get
\begin{equation}
\delta_2-\delta_1=\frac{\alpha Q^2_\psi}{4\pi}\int^1_0dx\left[-3(1-2x)\sum^\infty_{n=1}\log(n^2+\bar{c}^2)+6x(1-x)^2(mR)^2E^{\bar{c}^2}_1(1)\right]\, ,
\end{equation}
where $\bar{c}^2\equiv\bar{c}^2_{2F}=\bar{c}^2_{3V}$. Integrating by parts the first integral, we can see that the right-hand side of the above equation vanishes, showing in this way that $\delta_1=\delta_2$, as it happens in QED. The crucial step to prove this is:
\begin{equation}
\frac{d}{dx}\sum^\infty_{n=1}\log(n^2+\bar{c}^2)=E^{\bar{c}^2}_1(1)\frac{d\bar{c}^2}{dx}\, .
\end{equation}
The equation $\delta_1=\delta_2$ in turn implies, from (\ref{Rd}), that $Z_1=Z_2$ and then, from (\ref{Re}), that $e_B=Z^{-\frac{1}{2}}_3e$. Finally, from (\ref{FF1}), we can see that $F^{5D}_1(0)=1$.

\section{Vacuum polarization in $(4+n)$DQED}
\label{VPn} In subsection~\ref{PSE} we have seen that the ultraviolet divergence induced by the zero mode is canceled exactly by the ultraviolet divergences that arise from its KK excitations. This cancellation occurs as a result of the double multiplicity of KK excitations associated with zero mode [see Eq.(\ref{PDiv})]. In this section, we study the photon self-energy in QED with an arbitrary number $n$ of extra dimensions [$(4+n){\rm DQED}$] and show that such cancellation of ultraviolet divergences at one loop is exclusive of 5DQED.\\

\subsection{The renormalized vacuum polarization function}
In the UED approach, the $(4+n)$-dimensional high-energy theory is governed by the extended group ${\rm ISO}(1,3+n)\times{\rm U}_Q(1,{\cal M}^{4+n})$. In this case, the action of the theory is a functional of the ${\rm SO}(1,3+n)$ spinor $\Psi(x,\bar x)$ and the ${\rm SO}(1,3+n)$-vector gauge field ${\cal A}_M(x,\bar x)$. To describe physical effects at distance scales comparable to the size of the compact $n$-dimensional manifold, we need to implement the two canonical maps described in sections \ref{I} and \ref{5DQED}, obtaining in this way an effective theory governed by the standard group ${\rm ISO}(1,3)\times{\rm U}_Q(1,{\cal M}^{4})$. To simplify the analysis, we will assume an even number $n$ of extra dimensions, so there is chirality in this space. The spinor $\Psi(x,\bar x)$ of ${\rm SO}(1,3+n)$ is mapped into $2^{\frac{n}{2}}$ Dirac spinors $\psi(x,\bar x)$ of ${\rm SO}(1,3)$; while the ${\rm SO}(1,3+n)$ vector ${\cal A}_M(x,\bar x)$ is mapped into a vector ${\cal A}_\mu(x,\bar x)$ of ${\rm SO}(1,3)$ and $n$ scalars ${\cal A}_{\bar \mu}(x,\bar x)$ of ${\rm SO}(1,3)$ ($\bar \mu=5,\cdots, 4+n$). We assume that each coordinate $\bar x_i$ is coiled in a circle of radius $R_i$ and, as the case of one extra dimension, we introduce the orbifold $S^1/Z_2$. Then, in the case of $n$ extra dimensions we assume a compact manifold made of $n$ copies of the  $S^1/Z_2$ orbifold. It is assumed, for simplicity, that all radii are equal, that is, $R_1=\cdots=R_n\equiv R$. The compactified $(4+n){\rm DQED}$ theory is made of the standard fields $\psi^{(\underline{0})}(x)\equiv \psi(x)$ and $A^{(\underline{0})}_\mu(x)\equiv A_\mu(x)$, and their KK excitations, namely, $2^{\frac{n}{2}}$ KK Dirac spinors, $\psi^{(\underline{m})}(x)$, the gauge fields, $A^{(\underline{m})}_\mu(x)$, and their associated pseudo-Goldstone bosons $A^{(\underline{m})}_{\rm G}(x)$, as well as $n-1$ physical scalars $A^{(\underline{m})}_{\bar n}(x)$ ($\bar n=1,\cdots, n-1$). The masses of the KK spinors are given by $m^2_{\psi^{(\underline{m})}}=m^2+m^2_{(\underline{m})}$, while the masses of the KK gauge fields and scalar fields are given by $m^2_{(\underline{m})}\equiv p^{(\underline{m})}_{\bar \mu}p^{(\underline{m})}_{\bar \mu}$. Here $m^2_{(\underline{m})}=R^{-2}\underline{m}^2$, with  $\underline{m}^2=\underline{m}^2_1+\cdots+\underline{m}^2_n$ any admissible combination of Fourier indices.\\

The $(4+n){\rm DQED}$ theory generates many couplings among the diverse fields that compose it, but here we are interested only in those couplings that involve the electromagnetic field. These couplings are $A_\mu \bar \psi \psi$ and $A_\mu \bar {\psi}^{(\underline{m})} \psi^{(\underline{m})}$, whose Lorentz structure is dictated by the electromagnetic gauge group. The Feynman diagrams which contribute to the renormalized vacuum polarization function $\Pi_{(4+n){\rm D}}(q^2)$ are the same as those in Fig.~\ref{PS}, but now we have $2^{\frac{n}{2}}$ KK $\psi^{(\underline{m})}(x)$ fields circulating in the loop instead of the 2 spinor fields that characterize the 5DQED theory. Then, the one-loop contribution of $(4+n){\rm DQED}$ to the vacuum-polarization function can be written as follows:
\begin{equation}
\label{VPF1}
\Pi^{\rm loop}_{(4+n){\rm D}}(q^2)=-\frac{\alpha Q^2_{\psi}}{4\pi}\int^1_0 dx f_P(x)\Gamma\left(\frac{\epsilon}{2}\right)\left[\left(\frac{\Delta^2_{2P}}{4\pi \hat{\mu}^2}\right)^{-\frac{\epsilon}{2}}+2^{\frac{n}{2}}\sum_{(\underline{m})}\left(\frac{\Delta^2_{(\underline{m})P}}{4\pi \hat{\mu}^2}\right)^{-\frac{\epsilon}{2}}\right]\, ,
\end{equation}
where, by reasons that will be clear below, from now on we will use the symbol $\hat{\mu}$ instead of $\mu$ to denote the scale of the dimensional regularization scheme. In addition,  $\Delta^2_{(\underline{m})P}=m^2_{(\underline{m})}+\Delta^2_{2P}$. In the above expression, the symbol  $\sum_{(\underline{m})}$, $(\underline{m})\neq (\underline{0})$, summarizes a total of $ 2^{n}-1 $ different series and coincides with the notation $ \sum' $ used in Ref.~\cite{OP6}. In fact,
\begin{align}
\sum_{(\underline{m})}T^{(\underline{m})}  &:=  \sum_{m_{1}=1}^{\infty}T^{(m_{1},0,\ldots,0)}+\sum_{m_{2}=1}^{\infty}T^{(0,m_{2},0,\ldots,0)}+\ldots+ \sum_{m_{n}=1}^{\infty}T^{(0,\ldots,m_n)} \nonumber \\
&+\sum_{m_{1},m_{2}=1}^{\infty}T^{(m_{1},m_{2},0,\ldots,0)}+\ldots+ \sum_{m_{n-1},m_{n}=1}^{\infty}T^{(0,\ldots,0,m_{n-1},m_n)} \nonumber \\
&\vdots \nonumber \\
& +\sum_{m_{1},\ldots, m_{n}=1}^{\infty}T^{(m_{1},\ldots, m_{n})}\ .\label{SD}
\end{align}
Whereas positions of Fourier indices in the entries of $(\underline{m})$ are not relevant, the number of occupied entries makes a difference. So, in practice, one can use the following definition
\begin{equation}
\label{A2}
\sum_{(\underline{m})}=\sum^n_{l=1}\left(\begin{array}{ccc}
n \\
l
\end{array}\right)\sum^\infty_{m_1=1}\cdots \sum^\infty_{m_l=1}\, .
\end{equation}
Then, Eq.(\ref{VPF1}) can be written as
\begin{eqnarray}
\label{VPF2}
\Pi^{\rm loop}_{(4+n){\rm D}}(q^2)&=&-\frac{\alpha Q^2_{\psi}}{4\pi}\int^1_0 dx f_P(x)\Gamma\left(\frac{\epsilon}{2}\right)\Bigg[\left(\frac{\Delta^2_{2P}}{4\pi \hat{\mu}^2}\right)^{-\frac{\epsilon}{2}}+2^{\frac{n}{2}}\left(\frac{R^{-2}}{4\pi \hat{\mu}^2}\right)^{-\frac{\epsilon}{2}}\sum_{(\underline{m})}\left(\underline{m}^2+c^2_{2P}\right)^{-\frac{\epsilon}{2}}\Bigg]
\nonumber \\
&=&-\frac{\alpha Q^2_{\psi}}{4\pi}\int^1_0dx f_P(x)\left[\Gamma\left(\frac{\epsilon}{2}\right)\left(\frac{\Delta^2_{2P}}{4\pi \hat{\mu}^2}\right)^{-\frac{\epsilon}{2}}+2^{\frac{n}{2}}\left(\frac{R^{-2}}{4\pi \hat{\mu}^2}\right)^{-\frac{\epsilon}{2}}\sum^n_{l=1}\left(\begin{array}{ccc}
n \\
l
\end{array}\right)\Gamma\left(\frac{\epsilon}{2}\right)E^{c^2_{2P}}_l\left(\frac{\epsilon}{2}\right)\right]\, ,
\end{eqnarray}
where we have introduced the $l$-dimensional Epstein zeta function, which is defined as~\cite{K,E2,E3}
\begin{equation}
E^{c^2}_l(s)=\sum^{\infty}_{(m_1,\cdots,m_l)=1}\frac{1}{\left(m^2_1+\cdots +m^2_l+c^2_{2P}\right)^s}\, .
\end{equation}
Since the $l$-dimensional Epstein function $E^{c^2}_l(s)$ have poles at $s=\frac{l}{2}, \frac{l-1}{2},\cdots, -\frac{1}{2}, -\frac{3}{2}, \cdots$, except zero~\cite{K}, it is clear that  $E^{c^2}_l\left(\frac{\epsilon}{2}\right)$ converges for $\epsilon \to 0$. However, the way this happens is subtle, which can be appreciated more clearly by expressing the Epstein function in terms of the Riemann zeta function, whose properties are well known in the literature. One important result~\cite{E3} express the $l$-dimensional Epstein function in terms of the 1-dimensional one:
\begin{equation}
\label{EFR}
E^{c^2}_l(s)=\frac{(-1)^{l-1}}{2^{l-1}}\sum^{l-1}_{p=0}\left(\begin{array}{ccc}
l-1 \\
p
\end{array}\right)(-1)^p \pi^{\frac{p}{2}}\frac{\Gamma\left(s-\frac{p}{2}\right)}{\Gamma(s)}E^{c^2}_1\left(s-\frac{p}{2}\right)\, .
\end{equation}
The reduction to the Riemann zeta function is given through a power series in $c^2$, which, in our case, means to assume that $c^2_{2P}=\frac{\Delta^2_{2P}}{R^{-2}}< 1$, which is sufficient for our purposes, since our effective theory is, by definition, valid only for energies less than $R^{-1}$. The pass from $E^{c^2}_1(s)$ to $\zeta(s)$ is given by~\cite{ER}:
\begin{equation}
\label{EF1}
E^{c^2}_1(s)=\sum^\infty_{k=0}\frac{(-1)^k}{k!}\frac{\Gamma(k+s)}{\Gamma(s)}\zeta(2k+2s)c^{2k}\, ,
\end{equation}
where the Riemman function is defined by
\begin{equation}
\label{RF}
\zeta(s)=\sum^\infty_{n=1}\frac{1}{n^s}\, ,
\end{equation}
which has a simple pole at $s=1$. We can see from Eqs. (\ref{EFR}) and (\ref{EF1}) that finite terms which are the ratio of two quantities that diverge when $s$ tends to zero, as $ \frac{\zeta(1+s)}{\Gamma(s)}$, arise for $l>1$. This fact has nontrivial consequences when a product of the form $\Gamma(s)E^{c^2}_l(s)$, as the one appearing in Eq. (\ref{VPF2}), is considered. Thus, case $l> 1$ must be treated with some care. Using the above results, we can write,

\begin{equation}
\label{GR}
\sum^{n}_{l=1}\left(\begin{array}{ccc}
n \\
l
\end{array}\right)\Gamma\left(\frac{\epsilon}{2}\right)E^{c^2}_l\left(\frac{\epsilon}{2}\right)=
\frac{1}{2^{n-1}}\sum^n_{r=1}\sum^r_{l=1}\left(\begin{array}{ccc}
n \\
l-1
\end{array}\right)\pi^{\frac{n-r}{2}}\sum^\infty_{k=0}\frac{(-1)^k}{k!}
\Gamma\left(\frac{2k+r-n+\epsilon}{2}\right)\zeta(2k+r-n+\epsilon)c^{2k}\, .
\end{equation}
It is convenient to rewrite this expression so that its poles are evident. From the analytical properties of the gamma and Riemann functions, we can see that divergences arise for $2k+r-n$ an even integer less or qual to zero or for $2k+r-n=1$, which, for a given $n$, determine the poles of the $\Gamma(s)$ and $\zeta(s)$ functions, respectively. Then, after some rearrangements, Eq.~(\ref{GR}) can be written as
\begin{eqnarray}
\label{GR1}
\sum^{n}_{l=1}\left(\begin{array}{ccc}
n \\
l
\end{array}\right)\Gamma\left(\frac{\epsilon}{2}\right)E^{c^2}_l\left(\frac{\epsilon}{2}\right)&=&g_{(0)}(n)\,
\Gamma\left(\frac{\epsilon}{2}\right)\zeta(\epsilon)+F_{(0)}(n)+\sum^{[\frac{n}{2}]}_{k=1}\Big[f_{(k)}(n)\frac{1}{\sqrt{\pi}}
\Gamma\left(\frac{1+\epsilon}{2}\right)\zeta(1+\epsilon)\nonumber \\
&& +g_{(k)}(n)\Gamma\left(\frac{\epsilon}{2}\right)\zeta(\epsilon)+F_{(k)}(n)  \Big]c^{2k}+F(n,c^2)\, ,
\end{eqnarray}
where
\begin{subequations}
\begin{align}
g_{(0)}(n)&=2\left(1-\frac{1}{2^n}\right)\, , \\
F_{(0)}(n)&=\frac{1}{2^{n-1}}\sum^{n-1}_{r=1}\sum^{r}_{l=1}\left(\begin{array}{ccc}
n \\
l-1
\end{array}\right)\pi^{\frac{n-r}{2}}\Gamma\left(\frac{r-n}{2}\right)\zeta(r-n)\, , \, \, n>1\\
F_{(k)}(n)&=\frac{1}{2^{n-1}}\left(\sum^n_{r\neq n-(2k-1),n-2k} \right)\sum^r_{l=1}\left(\begin{array}{ccc}
n \\
l-1
\end{array}\right)\pi^{\frac{r-n}{2}}\Gamma\left(\frac{2k+r-n}{2}\right)\zeta(2k+r-n)\, , \, \, n>1\\
F(n,c^2)&=\frac{1}{2^{n-1}}\sum^n_{r=1}\sum^{r}_{l=1}\left(\begin{array}{ccc}
n \\
l-1
\end{array}\right)\pi^{\frac{n-r}{2}}\sum^{\infty}_{k=[\frac{n}{2}]+1}\frac{(-1)^k}{k!}\Gamma\left(\frac{2k+r-n}{2}\right)\zeta(2k+r-n)c^{2k}\, .
\end{align}
\end{subequations}
In the above expressions, the symbol $[\frac{n}{2}]$ means the floor of $\frac{n}{2}$, that is, the largest integer less than or equal to $\frac{n}{2}$. On the other hand, the $f_{(k)}(n)$ and $g_{(k)}(n)$ functions have the following property:
\begin{subequations}
\begin{align}
&f_{(k)}(1)=\cdots=f_{(k)}(2k-1)=0\, , \\
&g_{(k)}(1)=\cdots=g_{(k)}(2k)=0\, .
\end{align}
\end{subequations}
These relations implies that in the case of only one extra dimension, there are no divergences associated with the power series in $q^2$, since in this case $f_{(k)}(1)=0$ and $g_{(k)}(1)=0$ for all $k=1,2,\cdots$. In the case $n=2$, $g_{(k)}(2)=0$ for all $k$, but $f_{(1)}(2)\neq 0$ and $f_{(k)}(2)= 0$ for $k=2,3,\cdots$. If $n=3$, besides $f_{(1)}(3)\neq 0$, we have $g_{(1)}(3)\neq 0$, but $f_{(k)}(3)=0$ and $g_{(k)}(3)=0$ for all $k=2,3, \cdots$. Thus, for $n=2$ and $n=3$ divergences arise only in the first term ($k=1$) of the power series in $q^2$. However, for the cases $n=4$ and  $n=5$, divergences arise in both the first $(k=1)$ and second $(k=2)$ terms of the power series in $q^2$; $n=6$ and $n=7$ implies divergences in the terms $q^2$, $(q^2)^2$, and $(q^2)^3$, and os on. Explicit expressions of these functions are shown in Table~\ref{T1} for $k=1,2,3,4$.

\begin{table}[htbp]
\centering
\renewcommand{\arraystretch}{1.5}
\begin{tabular}{|c|c|c|c|c|c|c|}
\hline
$k$  & $f_{(k)}(n)$ & $g_{(k)}(n)$
\\ \hline
1& $-2\sqrt{\pi}\left(1-\frac{n+1}{2^n}\right)$& $-2\pi\left(1-\frac{n^2+n+2}{2^{n+1}}\right) $
\\ \hline
$2$ & $\pi^{\frac{3}{2}}\left(1-\frac{n^3+5n+6}{3\times 2^{n+1}}\right)$ & $\pi^2\left(1-\frac{n^4-2n^3+11n^2+14n+24}{3\times 2^{n+3}}\right)$
\\ \hline
$3$ & $-\frac{\pi^{\frac{5}{2}}}{3}\left(1-\frac{n^5-5n^4+25n^3+5n^2+94n+120}{15\times 2^{n+3}}\right)$ & $-\frac{\pi^3}{3}\left(1-\frac{n^6-9n^5+55n^4-75n^3+304n^2+444n+720}{45\times 2^{n+4}}\right)$
\\ \hline
$4$ & $\frac{\pi^{\frac{7}{2}}}{12}\left(1-\frac{n^7-14n^6+112n^5-350n^4+1099n^3+364n^2+3828n+5040}{315\times 2^{n+4}}\right)$ & $\frac{\pi^4}{12}\left(1-\frac{n^8-20n^7+210n^6-1064n^5+3969n^4-4340n^3+15980n^2+25584n+40320}{315\times 2^{n+7}}\right)$
\\ \hline
\end{tabular}
\caption{\label{T1} $f_{(k)}(n)$ and $g_{(k)}(n)$ as functions on the number $n$ of extra dimensions for the first four powers of $\mathbf{c}^2$.}
\end{table}
From the expression (\ref{GR1}), two types of divergences can be identified, one of which depends neither on the external moment nor on the compactification scale, which is characterized by the coefficient $g_{(0)(n)}$. Due to this, this type of divergences can be identified as short distances effects on the usual spacetime manifold ${\cal M}^4$, that is, they are usual ultraviolet divergences that are associated with divergences arising from the continuous sum $\int d^4k$ in the KK loop amplitudes. In contrast, the other type of divergence we have in (\ref{GR1}) depends on both the external momenta and the compactification scale. By its own nature, this type of divergence can be attributed to very high-energy effects or, equivalently, to short distances  effects on the compact manifold ${\cal N}^n$. To see this, note that the one-loop KK contributions behave like
\begin{equation}
\sum_{(\underline{k})}\int d^4k \frac{T_{\mu \nu \cdots}}{\left(k^2-\Delta^2_{(\underline{k})}\right)^m}=\sum_{(\underline{k})}\int d^4k \frac{T_{\mu \nu  \cdots}}{\left(k^2-k^2_{(\underline{k})}-\Delta^2_{(\underline{0})}\right)^m}\, ,
\end{equation}
where $k^2_{(\underline{k})}=m^2_{(\underline{k})}$ is the squared of the discrete momenta, that is, $k^2-k^2_{(\underline{k})}=k_Mk^M$, with $k_M=k_\mu +k_{\bar \mu}$. The right side of the above equation is quite suggestive, since it clearly shows us that divergences can arise either for very large $k_\mu$ momenta or for very large discrete $k_{\bar \mu}$ momenta, which may eventually result in a divergent continuous sum or in a divergent discrete sum, respectively. In fact, a very large $k^2_{(\underline{k})}$ implies a short distance effect in the compact manifold, since $1/k^2_{(\underline{k})}=R^{-2}/\underline{k}^2$ tends to zero for large combinations of Fourier indices $\underline{k}^2$. From these considerations, we can think of this new class of divergences as genuine ultraviolet divergences that can be handled by renormalization in a broader or modern sense, as is usually done in the context of the effective field theories approach, which are not renormalizable in the power counting sense~\cite{RMS}. Our general effective Lagrangian is given by Eq.~(\ref{ELR5}), but we only need to specify those interactions of canonical dimension higher than 4 needed to consistently remove divergences that emerge proportionally to powers of $(q^2/R^{2})$. Due to gauge invariance, the corresponding bare Lagrangian must be of the form
\begin{equation}
{\cal L}^{{\mathbf{d}>4}}_{B\, \gamma}=\sum^{[\frac{n}{2}]}_{k=1}\frac{\alpha_{(k)}}{(R^{-2})^k}\left(\partial_{\alpha_1}\cdots \partial_{\alpha_k}F_{\mu \nu}\right)
\left(\partial^{\alpha_1}\cdots \partial^{\alpha_k}F^{\mu \nu}\right)+\sum^{[\frac{n}{2}]}_{k=1}\frac{\delta^{(k)}_3}{(R^{-2})^k}\left(\partial_{\alpha_1}\cdots \partial_{\alpha_k}F_{\mu \nu}\right)
\left(\partial^{\alpha_1}\cdots \partial^{\alpha_k}F^{\mu \nu}\right)\, ,
\end{equation}
where $\delta^{(k)}_3\equiv Z_3 \alpha_{B(k)}-1$. Then, the renormalized polarization function can be written as follows:
\begin{equation}
\label{RPF}
\Pi_{(4+n)D}(q^2)=\sum^{[\frac{n}{2}]}_{k=1}\alpha_{(k)} \left(\frac{q^2}{R^{-2}}\right)^k+\Pi^{\rm loop}_{(4+n){\rm D}}(q^2)-\delta_3+
\sum^{[\frac{n}{2}]}_{k=1}\delta^{(k)}_3 \left(\frac{q^2}{R^{-2}}\right)^k \, .
\end{equation}
It is worth determining the counterterms using both a mass-independent scheme and a mass-dependent scheme. \\

\noindent \textit{Mass-independent scheme}. We use a $\overline{MS}$-like scheme, in which the counterterm is defined just to cancel the pole of the divergence plus some constant quantities that do not involve energy scales. From Eqs.~(\ref{VPF2}), (\ref{GR1}), and (\ref{RPF}), we can find the following counterterms:
\begin{subequations}
\begin{align}
\delta_3=&-\frac{\alpha Q^2_\psi}{4\pi }\left(\frac{4}{3}\right)\left[\frac{2}{\epsilon}-\gamma+\log(4\pi)+2^{\frac{n}{2}}g_{(0)}(n)\left[
-\frac{1}{\epsilon}-\frac{1}{2}\log\left(16\pi^3\right)+\frac{1}{2}\gamma\right] \right]\, , \\
\delta^{(k)}_3=&\frac{\alpha Q^2_{\psi}}{4\pi}2^{\frac{n}{2}}\left(\frac{4}{3}\right) 2^{\frac{n}{2}}\Bigg\{\left[f_{(k)}(n)-g_{(k)}(n)\right]\frac{1}{\epsilon}+f_{(k)}(n)\left[\gamma +\frac{1}{2}\log(4\pi)+\frac{1}{2}\psi^{(0)}\left(\frac{1}{2}\right)\right]\nonumber \\
&+g_{(k)}\left[\frac{1}{2}\gamma -\frac{1}{2}\log(16\pi^3)\right]\Bigg\}\, , \ \ \ \ k=1, \cdots, [\frac{n}{2}]\, ,
\end{align}
\end{subequations}
where $\psi^{(0)}\left(\frac{1}{2}\right)$ is the polygamma function of order zero and $\gamma$ is the Euler-Mascheroni constant. In this scheme, the renormalized polarization function is given by:
\begin{eqnarray}
\Pi_{(4+n)D}(q^2)&=&\frac{\alpha Q^2_\psi}{4\pi}\int^1_0 dx f_P(x)\Bigg\{\log\left(\frac{\Delta^2_{2P}}{\hat{\mu}^2}\right)-2^{\frac{n}{2}}\left[F(n,c^2_{2P})-g_{(0)}(n)\log\left(\frac{R^{-2}}{\hat{\mu}^2}\right) \right]\nonumber \\
&& +\sum^{[\frac{n}{2}]}_{k=1}\left[\alpha_{(k)}\left(\frac{q^2}{R^{-2}}\right)^k-
2^{\frac{n}{2}-1}\left(f_{(k)}(n)-g_{(k)}(n)\right)\log\left(\frac{R^{-2}}{\hat{\mu}^2}\right) \right]\Bigg\}
\end{eqnarray}
Note that this expression does not reduce to the usual one in the $R^{-1}\to \infty$ limit, which means that there is no decoupling of the new physics effects. This result is not surprising, since it is well known that in a mass-independent scheme the decoupling of heavy physics is not manifest. Next, we discuss a different renormalization scheme in which the de coupling of heavy physics is manifest.\\

\noindent \textit{Mass-dependent scheme}. The main feature of mass-independent schemes is that they do not involve a kinematical subtraction point. In contrast, mass-dependent schemes involve a subtraction point. Here, we choose an arbitrary subtraction point defined by $q^2=-\mu^2$, with $\mu$ the kinematical scale or subtraction scale. To determine the counterterms, we impose on the polarization function the following $[\frac{n}{2}]+1$  renormalization conditions:
\begin{subequations}
\begin{align}
&\Pi_{(4+n)D}(q^2=-\mu^2)=0 \, , \\
&\frac{d}{dq^2}\Pi_{(4+n)D}(q^2)\Big|_{q^2=-\mu^2}=0\, , \\
&\vdots \nonumber \\
&\frac{d^{[\frac{n}{2}]}}{d(q^2)^{[\frac{n}{2}]}}\Pi_{(4+n)D}(q^2)\Big|_{q^2=-\mu^2}=0\, .
\end{align}
\end{subequations}
To illustrate this renormalization scheme, we will study in detail the case $n = 2$. In such case, we have $[\frac{n}{2}]=1$, so from expression (\ref{RPF}) we have only the counterterms $\delta_3$ and $\delta^{(1)}_3$. A direct calculation leads to,
\begin{subequations}
\begin{align}
\label{D32}
\delta_3=&-\frac{\alpha Q^2_\psi}{4\pi}\int^1_0dxf_P(x)\Bigg\{\Gamma\left(\frac{\epsilon}{2}\right)\left(\frac{\bar{\Delta}^2_{2P}}{4\pi\hat{\mu}^2}\right)^{-\frac{\epsilon}{2}}
+\frac{x(1-x)\mu^2}{\bar{\Delta}^2_{2P}}\nonumber \\
&+2\left[A_{(0)}(2,\epsilon)+\left(\frac{m^2}{R^{-2}}\right)\hat{A}_{(1)}(2,\epsilon)\right] \nonumber \\
&+2\left[F(2,\bar{c}^2_{2P})+\left(\frac{m^2}{R^{-2}}-\bar{c}^2_{2P}\right)\frac{dF(2,\bar{c}^2_{2P})}{d\bar{c}^2_{2P}} \right]
\Bigg\}\, ,  \\
\label{DK32}
\delta^{(1)}_3=&-\alpha_{(1)}-A_{(1)}(2,\epsilon)+\frac{\alpha Q^2_\psi}{4\pi}\int^1_0dxf_P(x)\left[\frac{x(1-x)}{\bar{c}^2_{2P}}-2x(1-x)\frac{dF(2,\hat{c}^2_{2P})}{d\bar{c}^2_{2P}}\right]\, ,
\end{align}
\end{subequations}
where
\begin{subequations}
\begin{align}
A_{(0)}(2,\epsilon)=&\left(\frac{R^{-2}}{4\pi \hat{\mu}^2}\right)^{-\frac{\epsilon}{2}}g_{(0)}(2)\Gamma\left(\frac{\epsilon}{2}\right)\zeta(\epsilon)+F_{(0)}\, , \\
\hat{A}_{(1)}(2,\epsilon)=&\left(\frac{R^{-2}}{4\pi \hat{\mu}^2}\right)^{-\frac{\epsilon}{2}}f_{(1)}(2)\frac{1}{\sqrt{\pi}}\Gamma\left(\frac{1+\epsilon}{2}\right)\zeta(\epsilon)+F_{(1)}(2)\, ,\\
A_{(1)}(2,\epsilon)=&\frac{\alpha Q^2_\psi}{4\pi}\int^1_0 dx f_{P}(x)[x(1-x)]\hat{A}_{(1)}(2,\epsilon)
\end{align}
\end{subequations}
It is interesting to study more closely the divergent structure of the counterterms. Keeping only the poles of the gamma and zeta functions, we have for $\delta_3$
\begin{equation}
\delta_3=-\frac{\alpha Q^2_\psi}{4\pi}\left(\frac{4}{3}\right)\left\{\underbrace{\left(\frac{2}{\epsilon}\right)}_{SC}+
\underbrace{3\zeta(0)\left(\frac{2}{\epsilon}\right)-\sqrt{\pi}\left(\frac{m^2}{R^{-2}}\right)\left(\frac{1}{\epsilon}\right)}_{KKC}\right\}\, ,
\end{equation}
where the term divergent proportional to $(m^2/R^{-2})$ corresponds to a short distance effect in the compact manifold, which arises from the pole of the zeta function. The other two divergences correspond to short distances effects in the usual spacetime manifold. Note that the term $3\zeta(0)(2/\epsilon)=3\sum^\infty_{k=1}(2/\epsilon)=-(3/2)(2/\epsilon)$ represents the contribution to usual ultraviolet divergences of the infinite number of KK fields. It can be appreciated from this expression that the ultraviolet divergences induced by the KK excitations cannot cancel the one generated by the standard fermion for any $n=2$, as it occurs in the case $n=1$. It is easy to show that this true for all $n=2,4,\cdots $.\\

As far as the counterterm $\delta^{(1)}_3$ is concerned, its singular part can be written as follows:
\begin{equation}
\delta^{(1)}_3=-\frac{\alpha Q^2_\psi}{4\pi}\left(\frac{4}{3}\right)
\left\{\underbrace{\sqrt{\pi}\left(\frac{1}{\epsilon}\right)}_{KKC}\right\}\, ,
\end{equation}
which correspond to a short distance effect in the compact manifold because it arises from expression (\ref{GR1}) for the Epstein function given
as a power series in $c^2_{2P}$ or, equivalently, in powers of the external momenta.\\

Substituting Eqs.~(\ref{D32}) and (\ref{DK32}) into Eq.~(\ref{RPF}), we obtain the following renormalized polarization function:
\begin{eqnarray}
\label{RPFN2}
\Pi_{6D}(q^2)&=&-\frac{\alpha Q^2_\psi}{4\pi}\int^1_0dxf_P(x)\Bigg\{\log\left(\frac{\bar{\Delta}^2_{2P}}{\Delta^2_{2P}}\right)+\frac{x(1-x)q^2}{\bar{\Delta}^2_{2P}}\nonumber \\
&&+2\left[F(2,c^2_{2P})-F(2,\bar{c}^2_{2P})+(\bar{c}^2_{2P}-c^2_{2P})\frac{dF(2,\bar{c}^2_{2P})}{d\bar{c}^2_{2P}} \right] \Bigg\}\, .
\end{eqnarray}
In the above expressions, $\bar{c}^2_{2P}=c^2_{2P}|_{q^2=-\mu^2}=m^2+x(1-x)\mu^2$. Note that for $R^{-1}\to \infty$, the above expression reduces to the usual one obtained in this renormalization scheme, that is, the new physics effects decouple, as must be. This low energy behavior of the polarization function must be contrasted with that previously  found in a mass-independent scheme, in which there is no decoupling.\\

\subsection{The effective charge}
The vacuum polarization function allows us to define an \textit{effective charge}~\cite{EEC}:
\begin{equation}
\label{ECh}
\alpha_{\rm eff}(q^2)=\frac{e^2_B}{4\pi}\frac{1}{1-\Pi_B(q^2)}=\frac{\alpha}{1-\Pi(q^2)}\, ,
\end{equation}
where $\Pi_B(q^2)$ and $\Pi(q^2)$ are the bare and renormalized vacuum polarization functions, respectively. The effective charge has the following properties: (1) it is gauge independent, since the vacuum polarization function is gauge independent to all orders; (2) as a consequence of the Ward identity discussed in subsection \ref{WI}, the effective charge can be expressed in terms of bare quantities, so it is both renormalization scale- and scheme-independent; (3) at $q^2=0$, it matches the fine structure constant $\alpha_{\rm eff}(0)=\alpha=\frac{1}{137.035\cdots}$; (4) for $\frac{-q^2}{m^2}\ll 1$, it gives the correction to the Coulomb's law for the interaction between two static heavy charges; (5) the virtual contribution of a fermion $f$ to the renormalized one-loop vacuum polarization can be reconstructed directly from the tree-level cross section $\sigma(\psi^- \psi^+ \to f^-f^+)$. The last two points arise as a direct consequence of the analytical properties of the polarization function. It is important to mention that this is no longer true when QED is embedded in the electroweak sector of the SM, as the $W$ gauge boson contribution leads to a vacuum polarization which is gauge dependent\footnote{The possibility of extending the QED concept of effective charge to the non-abelian case has been studied~\cite{GIEC} in the context of the Pinch Technique~\cite{PT}.}.\\

In our case, $(4+n){\rm DQED}$ predicts, at the one-loop level, an effective charge given by
\begin{equation}
\label{EChED}
\alpha^{(4+n){\rm D}}_{\rm eff}(q^2)=\frac{\alpha}{1-\Pi_{(4+n)D}(q^2)}\, ,
\end{equation}
with $\Pi_{(4+n){\rm D}}(q^2)$ given by Eq.(\ref{RPF}). It should be noted that the KK contribution has the same analytical structure as the QED contribution. Because of this, one expects $\alpha^{(4+n){\rm D}}_{\rm eff}(q^2)$ to possess the same properties as $\alpha_{\rm eff}(q^2)$. In fact, we can see that properties (1) to (3) are clearly fulfilled. We now proceed to analyze the modifications introduced by extra dimensions on the properties (4) and (5). To simplify the analysis, we will consider only one extra dimension.\\

\begin{figure}
\centering
\includegraphics[width=14cm]{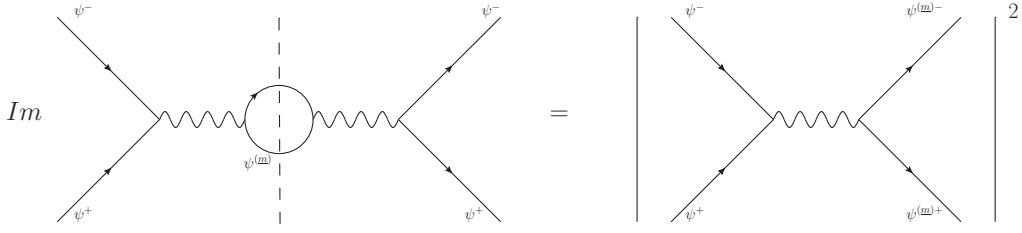}
\caption{\label{CS} {\footnotesize The relation between the imaginary part of the $\psi^{(n)}$ contribution to the one-loop vacuum polarization
and the tree-level cross section $\sigma\left(\bar{\psi} \psi \to \bar{\psi}^{(n)}\psi^{(n)}\right)$ in 5DQED.}}
\end{figure}

 We begin by studying the analytical structure of $\Pi_{{\rm 5D}}(q^2)$. We note that for $q^2<0$, which corresponds to the $t$- or $u$-channel, $\Pi_{{\rm 5D}}(q^2)$ is real and well defined. However, when $q^2>0$, which corresponds to the $s$-channel, $\Pi_{{\rm 5D}}(q^2)$ can have an imaginary part. In this case, the logarithms that define $\Pi_{\rm 5D}(q^2)$ can have branch cuts when their arguments are negative. Since the $x(1-x)$ factor is at most $1/4$, the logarithms in $\Pi_{\rm 5D}(q^2)$ have branch cuts beginning at
 \begin{equation}
 q^2=4m^2, \, \ \ q^2=4m^2_{\psi^{(1)}}\, , \ \  q^2=4m^2_{\psi^{(2)}}\, , \cdots \, .
 \cdots\, ,
 \end{equation}
 Assume that $m^2_{\psi^{(n)}}-x(1-x)q^2$ is negative up to a given Fourier index $N$, so $\Pi_{\rm 5D}(q^2)$ have negative logarithms up to the term $m^2_{\psi^{(N)}}-x(1-x)q^2$. This means that we have $N$ replicas of the QED case. The QED result is well known in the literature (see, for instance, Ref.~\cite{Peskin}), so we will just present the prediction of 5DQED,
\begin{equation}
\label{IMP}
{\rm Im} \left[ \Pi_{\rm 5D}(q^2)\right]=\frac{\alpha Q^2_{\psi}}{3}\left[\beta\left(1+\frac{2m^2}{q^2}\right)+2\sum^N_{n=1}\beta_{(n)}\left(1+\frac{2m^2_{\psi^{(n)}}}{q^2}\right)\right]\, ,
 \end{equation}
 where $\beta=\sqrt{1-4m^2/q^2}$ and  $\beta_{(n)}=\sqrt{1-4m^2_{\psi^{(n)}}/q^2}$. The factor of 2 multiplying the KK contribution in (\ref{IMP}) is due to the two degenerate excitations $\psi^{(n)}_{(1)}$ and $\psi^{(n)}_{(2)}$. Eq.~(\ref{IMP}) implies the well-known fact that the one-loop contribution of the $\psi^{(n)}$ fermion to the vacuum polarization determines the cross sector of the $\bar{\psi} \psi \to \bar{\psi}^{(n)}\psi^{(n)}$ process. Such a connection is given by:
 \begin{equation}
 \frac{e^2}{q^2}{\rm Im}\left[ \Pi_{5D}(q^2)\right]_{\beta_{(n)}}=\sigma\left(\bar{\psi} \psi \to \bar{\psi}^{(n)}\psi^{(n)}\right)\, ,
 \end{equation}
where the subscript ${\beta_{(n)}}$ in ${\rm Im}\left[ \Pi_{5D}(q^2)\right]$ indicates that only the term proportional to ${\beta_{(n)}}$ is considered, as it is shown in Fig.~\ref{CS}.\\

The imaginary part of $\Pi_{\rm 5D}(q^2)$ also induces a correction on the Coulomb's Law, that is, $V_{\rm eff}(r)=V(r)+\delta V(r)$, where $V(r)$ is the classical potential and $\delta V(r)$ the quantum correction. Following Ref.~\cite{Peskin}, we obtain in our case:
\begin{equation}
\delta V_{\rm 5D}(r)\approx-\frac{\alpha^2}{4\sqrt{\pi}\, r} \frac{e^{-2mr}}{(mr)^{\frac{3}{2}}}\left[1+2\sum^N_{n=1}\frac{e^{-2n\left(\frac{r}{R}\right)}}{\left(1+\frac{n}{mR}\right)^{\frac{3}{2}}} \right]\, .
\end{equation}
Note that the QED result is recovered in the limit as $R\to 0$.

\subsection{The beta function}
The beta function measures the variation of the coupling constant with energy. It is defined by
\begin{equation}
\beta(e)=\mu\frac{d e}{d \mu}\, ,
\end{equation}
where $\mu$ is an energy scale. The purpose of this section is to calculate the beta function in the context of $(4+n){\rm DQED}$. In theories in which the particle masses can be ignored, the calculation of the beta function is actually very simple if a mass-independent subtraction scheme is used, such as MS or $\overline{\rm MS}$, which do not involve a kinematical point to define the counterterm. MS-like schemes have an disadvantage, because heavy particles do not decouple at energies below their masses, as required by the decoupling theorem~\cite{AC}. Since the beta function is a physical quantity, effects of extra dimensions must decouple at energies much smaller than the compactification scale $R^{-1}$. In our case, it is clear that we should not use a MS-like scheme because the KK mass spectrum comprises a wide range of energy. Due to this, we will compute the beta function using a mass-dependent scheme. In particular, we will use the $\mu$-scheme already introduced in this section. It is worth studying the beta function in cases $n = 1$ and $n = 2$.\\

\noindent \textit{The case $n=1$}. Using the renormalization condition
\begin{equation}
\Pi_{5D}(q^2=-\mu^2)=0\, ,
\end{equation}
we obtain the following counterterm
\begin{equation}
\delta_3=-\frac{\alpha}{4\pi}\int^1_0dxf_P(x)\left[
\Gamma\left(\frac{\epsilon}{2}\right)\left(\frac{\bar{\Delta}^2_{2P}}{4\pi\hat{\mu}^2}\right)^{-\frac{\epsilon}{2}}+
2\left(\frac{R^{-2}}{4\pi\hat{\mu}^2}\right)^{-\frac{\epsilon}{2}}\Gamma\left(\frac{\epsilon}{2}\right)E^{\bar{c}^2_{2P}}_1(1)\right]\, ,
\end{equation}
where we have put $Q^2_\psi=1$ for simplicity. This counterterm leads to the following beta function:
\begin{eqnarray}
\beta_{5D}(e)&=&e\mu^2 \frac{\partial \delta_3}{\partial \mu^2} \nonumber \\
&=&\frac{e^3}{(4\pi)^2}\int^1_0dx f_P(x)\left[\frac{x(1-x)\mu^2}{m^2+x(1-x)\mu^2}+2\sum^\infty_{k=1}\frac{x(1-x)\mu^2}{m^2_{\psi^{(k)}}+x(1-x)\mu^2}\right]\nonumber \\
&=&\frac{e^3}{(4\pi)^2}\int^1_0dx f_P(x)\left[\frac{x(1-x)\mu^2}{m^2+x(1-x)\mu^2}+\frac{\sqrt{x(1-x)}\mu}{R^{-1}}\coth\left(\frac{\sqrt{x(1-x)}\mu}{R^{-1}}\right)-1 \right]\, .
\end{eqnarray}
This beta function depends on the ratio of fermion masses $m$, $m^2_{\psi^{(\underline{m})}}$ and the subtraction point $\mu$. In particular, when $m\ll \mu$ the QED contribution approaches the value obtained in the $\overline{\rm MS}$ scheme, so
\begin{equation}
\label{B2}
\beta(e)=\frac{e^3}{12\pi^2}+\frac{e^3 }{(4\pi)^2}\int^1_0 f_P(x)\left[\frac{\sqrt{x(1-x)}\mu}{R^{-1}}\coth\left(\frac{\sqrt{x(1-x)}\mu}{R^{-1}}\right)-1\right]\, .
\end{equation}
If on the other hand, $m\ll \mu \ll R^{-1}$, we recover the QED result
\begin{equation}
\label{B3}
\beta(e)=\frac{e^3}{12\pi^2}\, .
\end{equation}
 This result shows us that the new physics effects are of decoupling nature. Another interesting scenario is when $m\gg \mu$, in which the beta function vanishes.\\

\noindent \textit{The case $n=2$}. In this case, there are two coupling constants, namely $ e $ and $\alpha_{(1)}$, each with its associated beta function, but we will limit ourselves to analyzing the usual beta function. From Eq. (\ref{D32}), a direct calculation leads to
\begin{equation}
\beta_{6D}(e)=\frac{e^3}{(4\pi)^2}\int^1_0dxf_P(x)\left\{\frac{\left[x(1-x)\mu^2\right]^2}{\left[m^2+x(1-x)\mu^2\right]^2}+
2x(1-x)\left(\frac{\mu^2}{R^{-2}}\right)\bar{c}^2_{2P}\frac{d^2F(2,\bar{c}^2_{2P}}{d(\bar{c}^2_{2P})^2}\right\}\, .
\end{equation}
This expression can be rewritten in terms of Epstein functions using the following identity
\begin{equation}
\frac{d^2F(2,\bar{c}^2_{2P})}{d(\bar{c}^2_{2P})^2}=2E^{\bar{c}^2_{2P}}_1(2)+E^{\bar{c}^2_{2P}}_2(2)\, ,
\end{equation}
which arises from Eq.~(\ref{GR}). Note that both $E^{\bar{c}^2_{2P}}_1(2)$ and $E^{\bar{c}^2_{2P}}_2(2)$ Epstein functions are convergent. Introducing these functions, one has
\begin{equation}
\beta_{6D}(e)=\frac{e^3}{(4\pi)^2}\int^1_0dxf_P(x)\left\{\frac{\left[x(1-x)\mu^2\right]^2}{\left[m^2+x(1-x)\mu^2\right]^2}+
2x(1-x)\left(\frac{\mu^2}{R^{-2}}\right)\bar{c}^2_{2P}\left[2E^{\bar{c}^2_{2P}}_1(2)+E^{\bar{c}^2_{2P}}_2(2)\right]\right\}\, .
\end{equation}
Thus, for $m\ll \mu$, we have the usual beta function obtained in the $\overline{MS}$ scheme plus a new physics correction given by a power series in $\mu/R^{-1}$,
\begin{equation}
\beta_{6D}(e)=\frac{e^3}{12\pi^2}+\frac{e^3}{(4\pi)^2}\int^1_0dxf_P(x) 2(\bar{c}^2_{2P})^2\left[2E^{\bar{c}^2_{2P}}_1(2)+E^{\bar{c}^2_{2P}}_2(2)\right]\, ,
\end{equation}
which is a behavior similar to the case of only one extra dimension. The scenarios $m\ll \mu \ll R^{-1}$ and $m\gg \mu$ are also identical to the case of only one extra dimension.

\section{Summary}
\label{C} In this paper, we have comprehensively studied the one-loop structure of the fermion self-energy, the photon self-energy, and the vertex function in QED with one extra dimension. The discrete and continuous sums that characterize the one-loop amplitudes in this type of theories were regularized using the dimensional-regularization scheme. As a consequence, the KK contribution to these one-loop amplitudes is proportional to products of the gamma function and the Epstein function, both depending on the complex number $\epsilon=4-D$. Such contributions are proportional to $\Gamma(\epsilon/2)E^{c^2}_1(\epsilon/2)$. This expression tells us that the role of the $ E^{c^2}_1 (\epsilon / 2) $ function is to quantify the impact of the ultraviolet divergences induced by the infinite number of KK fields. However, the 1-dimensional Epstein function $E^{c^2}_1(s)$ has poles at $s=1/2, -1/2,-3/2,\cdots$, so this infinite sum is convergent in the $\epsilon \to 0$ limit. Consequently,   $\Gamma(\epsilon/2)E^{c^2}_1(\epsilon/2)=\Gamma(\epsilon/2)E^{c^2}_1(0)=\Gamma(\epsilon/2)\zeta(0)$, with the Riemann zeta function having the value $\zeta(0)=\sum^\infty_{n=1}=-\frac{1}{2}$. Due to this property of the Epstein function, we were able to define renormalized quantities that are reduced to the usual ones of QED in the $R^{-1}\to \infty$ limit. It was shown that 5DQED fulfills the Ward identity satisfied by QED, which in turn implies that $e_B=Z^{-\frac{1}{2}}_3e$, being this the main feature of abelian gauge theories. In the case of the photon self-energy, 5-dimensional QED has a double multiplicity of KK fermions, so their infinite number of excitations induce an ultraviolet divergence proportional to $2\zeta(0)=-1$, which cancels the ultraviolet divergence generated by the zero mode. Nonetheless, this curious result is exclusive of QED with only 1 extra dimension. The correction induced by the extra dimension on the anomalous magnetic dipole moment was calculated, showing that it is free of both infrared and ultraviolet divergences and reduces to the usual QED result in the $ R^{-1} \to \infty $ limit. Since any vertex function of canonical dimension higher than 4 is proportional to $\Gamma(N+\epsilon/2)E^{c^2}_1(N+\epsilon/2)$, with $N$ an integer number, it is clear that this type of products is free of divergences, so, at the one-loop level, the only ultraviolet divergences in 5DQED are the usual ones of QED.\\

The photon self-energy was also explored in the context of QED with an arbitrary even number $n$ of extra dimensions. It was shown that for $n\geq 2$, two types of divergences emerges from the Epstein functions appearing in the loop amplitudes through the sum of products $\sum^{n}_{l=1}\left(\begin{array}{ccc}
n \\
l
\end{array}\right)\Gamma\left(\frac{\epsilon}{2}\right)E^{c^2}_l\left(\frac{\epsilon}{2}\right)$. These types of divergences arise as a consequence of reducing multidimensional Epstein functions into the one-dimensional one. For example, a two-dimensional function breaks down as $E^{c^2}_2\left(\frac{\epsilon}{2}\right)=-\frac{1}{2}E^{c^2}_1\left(\frac{\epsilon}{2}\right)+
\frac{\sqrt{\pi}}{2}\Gamma\left(-\frac{1}{2}+
\frac{\epsilon}{2}\right)\frac{E^{c^2}_1\left(-\frac{1}{2}+\frac{\epsilon}{2}\right)}{\Gamma\left(\frac{\epsilon}{2}\right)}$. In the last term, $E^{c^2}_1\left(-\frac{1}{2}+\frac{\epsilon}{2}\right)$ diverges for $\epsilon \to 0$ but the pole is canceled by the gamma function, so $E^{c^2}_2\left(\frac{\epsilon}{2}\right)$ converges in this limit. However, when we consider the product $ \Gamma\left(\frac{\epsilon}{2}\right)E^{c^2}_2\left(\frac{\epsilon}{2}\right)$,  the result is the presence of two types of divergences, one type associated with the pole of the gamma function and the other from the pole of the one-dimensional Epstein function. Since the one-dimensional Epstein function is in turn expressed as a power series in $c^2\sim \frac{q^2}{R^{-2}}$, the divergences associated to it emerge naturally as coefficients of powers of the external momenta. We argued that the first type of divergences, which do not involve the compactification scale, are usual ultraviolet divergences in the sense that they correspond to short distance effects in the usual spacetime manifold. On the other hand, the second type of divergences, which  appear as coefficients of powers of $\frac{q^2}{R^{-2}}$, are also genuine ultraviolet divergences since they arise from short distance effects in the compact manifold, so they can be removed from amplitudes through renormalization. To generate the required counterterms, interactions of canonical dimension higher than four must be introduced, which is not an obstacle, since such interactions are already available in our effective Lagrangian, which contains all interactions compatible with the ${\rm ISO}(1,3)\times U_Q(1,{\cal M}^4)$ symmetry. The way to implement renormalization in a modern or broader sense in Kaluza-Klein theories was studied in detail. The vacuum polarization function $\Pi_{(4+n){\rm D}}(q^2)$ was calculated and used to study the properties of the effective charge. Since the KK contributions to the polarization function have the same analytical structure as the zero mode contribution, the main properties of the effective charge are automatically fulfilled. By way of illustration, the relation between the imaginary part of the $\psi^{(n)}$ contribution to the one-loop vacuum polarization and its relation with the tree-level cross section $\sigma\left(\bar{\psi} \psi \to \bar{\psi}^{(n)}\psi^{(n)}\right)$ was discussed in 5DQED. The contribution from the quantum correction to Coulomb's Law was calculated in 5DQED.\\

One important contribution of our work is the study of the beta function in the cases $5{\rm DQED}$ and $6{\rm DQED}$. The calculation of this function was performed using a mass-dependent scheme, instead of mass-independent schemes that are commonly used in QED, as MS-like schemes. The reason for this is that KK theories have a mass spectrum that covers a wide range of energies. Because of this, we use the so called $\mu$-scheme, with $\mu$ an arbitrary subtraction point. The beta function so obtained fulfills all physical requirements. In particular, for energies $m\ll \mu \ll R^{-1}$ it reduces to the well-known result of QED obtained in a mass-independent scheme. For $m\gg \mu$, both the zero mode and KK excitations decouple, so the beta function vanishes, as it occurs in QED. All these facts clearly reflect that our beta function fulfills all desirable physical requirements.\\

\acknowledgments{We acknowledge financial support from CONACYT and
SNI (M\' exico).}

\end{document}